\documentclass[epj-spec]{svjour}
\usepackage{amsmath}
\usepackage{graphics}
\begin{document}

\title{Nonlinear optical effects in artificial materials}

\author{ Andrei I. Maimistov\inst{1}\fnmsep
\thanks{\email{maimistov@pico.mephi.ru}} \and Ildar R.
Gabitov\inst{2,3}\fnmsep\thanks{\email{gabitov@math.arizona.edu}} }
\institute{Department of Solid State Physics, Moscow Engineering
Physics Institute, \\Kashirskoe sh. 31, Moscow, 115409 Russia \and
Department of Mathematics, University of Arizona, \\ 617 North Santa
Rita Avenue, Tucson, Arizona 85721, USA \and L.D. Landau Institute
for Theoretical Physics, \\2 Kosygin St., Moscow 119334, Russia}

\abstract{ We consider some nonlinear phenomena in metamaterials
with negative refractive index properties. Our consideration
includes a survey of previously known results as well as
identification of the phenomena that are important for applications
of this new field. We focus on optical behavior of thin films as
well as  multi-wave interactions. }

\maketitle

\section{Introduction}

\label{intro}

In recent years, the development of nanotechnology has led to the
creation of new materials  with very unusual optical properties -
metamaterials. Metamaterials are composed of host dielectric
materials with embedded periodic structures, patterned on a scale
much shorter than the operating wavelength and thus on a nanometer
scale if the infrared or visible spectra are targeted. Though they
are made of positive index materials at small length scale, these
structures exhibit abnormal dispersion characteristics at optical
scales. The material can behave as a medium with \emph{negative
index of refraction} associated with negative values of the
effective permittivity $\varepsilon $ and effective permeability
$\mu $~\cite{SSS01}.  The electric field, magnetic field wavevector
then constitute a left handed system, where the wavevector and the
Poynting vector are antiparallel~\cite{M45,V68,P100,P200}. In this
case, they are termed \emph{left handed materials }(LHMs). Beyond
this, inversion of the phase velocity with respect to the direction
of propagation of energy leads to negative refraction. That means
both the incident and refracted beam are on the same side of the
normal to the refraction interface, as if predicted by Snell's law
with a negative refractive index.

Left handed properties of metamaterials are most pronounced in two
situations. The first is in the behavior of  a quasi-monochromatic
wave at the interface of a left-handed and a conventional right
handed material. In particular we consider layered materials because
the use of thin films of LHM opens new opportunities for design of
photonic crystals with unusual properties. The second is the
interaction of two waves with different carrier frequencies which
experience different signs of refractive index. Such a difference is
a consequence of how   LHM are realized using simultaneous resonance
of electric and magnetic field components with metallic
nanostructures embedded in a host dielectric. Such a material
behaves as right handed for  waves with carrier frequencies above
the resonance region and as left-handed for  waves with carrier
frequencies below the resonance region.

In this paper we consider only phenomena related to these two
cases.

\section{ Nonlinear waveguiding phenomena on the interface of
  materials with opposite signs of their refractive indices}

 \label{sec:1}

 Let us consider a  layer of left-handed material   on a dielectric
 substrate with positive refraction index - a right handed material
 (RHM). We  study optical waves propagating along this interface in the
 $z$-direction of the Cartesian frame. The normal to the surface
 is chosen to be  the $x$- direction while the $y$-axis lies in the
 interface plane. There are two types of the surface waves: (a)
 transverse electric waves (TE waves) with $\mathbf{E}=(0.E_{y},0)$
 and $\mathbf{H} =(H_{x},0,H_{z})$, and (b) transverse magnetic
 waves (TM waves) with $ \mathbf{E}=(E_{x},0,E_{z})$\ and
 $\mathbf{H}=(0.H_{y},0)$. If the tensor of dielectric permeability
 of one of the media is a square-law function of electrical field
 strength \ $\hat{\varepsilon}(\omega ,\mathbf{E})=\hat{
 \varepsilon}(\omega )+\hat{\varepsilon}_{nl}(\omega ):\mathbf{EE}$\
 \ then propagation of the non-linear surface polaritons (NLSP)  is
 possible for both the TM and the TE waves~\cite{Darman,Ruppin2,Shadr4}.
 Linear surface polaritons corresponding to
 $\hat{\varepsilon} _{nl}(\omega )=0$ were found in~\cite{Ruppin1}.
 Without loss of generality, here we will consider only  TE waves.

 \subsection{\ Non-linear surface waves of TE-type}

 Let us consider waves on the  interface between two isotropic media.
 The electrodynamic properties of each medium  are described by effective
 dielectric permittivity and effective magnetic permeability. In the
 linear limit at $x<0$ the medium is characterized by positive
 permittivity and permeability $\varepsilon _{1}>0,\mu _{1}>0$ and at
 $x>0$ both $ \varepsilon $ and $\mu _{2}<0$ are chosen to be
 negative $ \varepsilon _{2}<0,~~\mu _{2}<0$. Due to its nature the
 electrical field vector of the TE-waves has only one nonzero
 component, $E_{y}=E$, which is governed by the equation:
 \begin{equation}
 E_{,zz}+E_{,xx}+k_{0}^{2}\varepsilon (\omega ,\mathbf{E})\mu (\omega
 , \mathbf{E})E=0,  \label{eq812}
 \end{equation}
 where $k_{0}=\omega /c$. The magnetic field  vector components  are
 coupled with $E$ by the following relations:
 \begin{equation*}
 H_{x}=i(k_{0}\mu )^{-1}E_{,z},\quad H_{z}=-i(k_{0}\mu )^{-1}E_{,x}
 \end{equation*}
 On the interface plane at $x=0$ the field components $E$, $H_{x}$
 and $H_{z}$ satisfy  the set of jump conditions:
 \begin{equation*}
 E(0-)=E(0+),~~ H_{z}(0-)=H_{z}(0+),\quad \mu _{1}H_{x}(0-)=\mu _{2}H_{z}(0+).
 \end{equation*}
 Due to the translation symmetry along the $z$-axis, the $y$-component of
 the electric field $E(\omega ,x,z)$ can be chosen as
 \begin{equation}
 E(\omega ,x,z)=\Phi (\omega ,x)\exp [i\beta (\omega )z],
 \label{eq814}
 \end{equation}
 where $\beta (\omega )$ is a propagation constant \cite{Integ}.
 The equation for the transverse profile of the electrical field of NLSW
 follows from equations (\ref{eq812}) and (\ref{eq814}):
 \begin{equation}
 \Phi _{,xx}+[k_{0}^{2}\varepsilon (\omega ,x)\mu (\omega ,x)-\beta ^{2}]\Phi
=0.  \label{eq815}
\end{equation}
The above jump conditions now can be represented in terms of  the
amplitude function $\Phi (x)$:
\begin{equation}
\Phi (0-)=\Phi (0+),\quad \mu _{1}^{-1}\Phi _{,x}(0-)=\mu
_{2}^{-1}\Phi _{,x}(0+),  \label{eqA12}
\end{equation}
The boundary conditions
\begin{equation}
\underset{|x|\rightarrow \infty }{\lim }\Phi (\omega ,x)=0,\quad
\underset{ |x|\rightarrow \infty }{\lim }\Phi _{,x}(\omega ,x)=0
\label{eqA13}
\end{equation}
select the   surface wave  solutions of~(\ref{eq815}).

As an example giving rise to nonlinear polaritons we considered the
simple geometry where the  nonlinear left-handed material  is placed
on a linear semi-infinite conventional right-handed material:
\begin{equation*}
\varepsilon (\omega ,x)=\left\{
\begin{array}{cc}
\varepsilon _{1}, & x<0 \\
\varepsilon _{2}+\varepsilon _{nl}\Phi ^{2}, & x>0
\end{array}
\right. ,\quad \mu (\omega ,x)=\left\{
\begin{array}{cc}
\mu _{1}, & x<0 \\
\mu _{2}, & x>0
\end{array}
\right.
\end{equation*}
Equation (\ref{eq815}) can be solved at $x<0$\ and at $x>0$, where
the media are homogeneous,  taking account of the boundary
conditions (\ref{eqA13}). The matching condition for  these
solutions at the boundary $x=0$ results in dispersion relations for
NLSP.

The equation (\ref{eq815} ) does not have bounded solutions, if $\mu
_{2}\varepsilon _{nl}<0$ (self-refocusing LHM). This means that
nonlinear surface polaritons do not exist in this geometry with
cubic nonlinearity in presence of  a self-refocusing left-handed
material.  It can be shown that, if $\mu _{2}\varepsilon _{nl}>0$
(this is the case of a self- focusing non-linear LHM), then NLSP
exists only when $p^{2}=(\beta ^{2}-k_{0}^{2}\varepsilon _{2}\mu
_{2})>0$. In this case the solution of equation (\ref{eq815}) reads
\begin{equation}
\Phi (x)=\left\{
\begin{array}{cc}
A_{1}\exp (qx), & x<0 \\
\alpha ^{-1/2}\sec \mathrm{h}[p(x-x_{2})], & x>0,
\end{array}
\right.  \label{eqA8110}
\end{equation}
where $\alpha =k_{0}^{2}\varepsilon _{2}\mu _{2}/2p^{2}$,
$q^{2}=(\beta ^{2}-k_{0}^{2}\varepsilon _{1}\mu _{1})>0$, and
$x_{2}$ is the location of the maximum of the electrical field.  The
continuity condition at the interface $x=0$ requires
\begin{equation}
A_{1}=\alpha ^{-1/2}~\sec \mathrm{h}(px_{2}),  \label{eqA8113a}
\end{equation}
\begin{equation}
\mu _{2}q=\mu _{1}p~\tanh (px_{2}).  \label{eqA8113b}
\end{equation}
The expression (\ref{eqA8113a}) defines $x_2$ as function of
electric field strength $A_{1}$ at the interface. This equation
(\ref{eqA8113b}) is the dispersion relation for NLSP. Since $mu
_{1}$  and $\mu _{2}$ have different signs ( $\mu _{1}>0$, $\mu
_{2}<0$), the parameter $x_{2}$ must be negative. Thus the electric
field reaches its   maximum value on the interface  at $x=0$. It
should be noted that in the case where both linear and nonlinear
dielectrics are right-handed,  the electric field reaches its
maximum value inside the non-linear dielectric medium, i.e., where
$x_{2}$ is positive.

In the other  case, when a linear LHM is placed on a nonlinear
semi-infinite medium  (RHM), $\varepsilon$ and $\mu$ are defined as
follows:
\begin{equation*}
\varepsilon (\omega ,x)=\left\{
\begin{array}{cc}
\varepsilon _{1}+\varepsilon _{nl}\Phi ^{2}, & x<0 \\
\varepsilon _{2}, & x>0
\end{array}
\right. ,\quad \mu (\omega ,x)=\left\{
\begin{array}{cc}
\mu _{1}, & x<0 \\
\mu _{2}, & x>0
\end{array}
\right.
\end{equation*}
The governing equations and corresponding solutions in this case can
be easily obtained from equations~(\ref{eqA8110})
and~(\ref{eqA8113a}),~(\ref{eqA8113b}).

For a self-refocusing medium, solutions of  equation (\ref{eq815})
are given by the expression
\begin{equation}
\Phi (x)=\left\{
\begin{array}{cc}
\pm |\alpha |^{-1/2}\mathrm{co}\sec \mathrm{h}[q(x-x_{1})], & x<0 \\
A_{2}\exp (-px), & x>0
\end{array}
\right. .  \label{eq8112}
\end{equation}
The continuity condition   at the interface $x=0$ requires
\begin{equation}
A_{2}=\pm |\alpha |^{-1/2}/\sinh (-qx_{1})  \label{eqA811da}
\end{equation}
\begin{equation}
\mu _{1}q=\mu _{2}q\tanh (qx_{1})  \label{eqA811db}
\end{equation}
Since $ \mu _{1}>0$ and $\mu _{2}<0$, it follows from the dispersion
relation (\ref{eqA811db})  that the parameter $x_1$ must be
negative. Thus the electric field reaches its maximum value on the
interface $x=0$ as in  the case of a self-refocusing medium.

A detailed analysis of   the surface waves on the interface between
nonlinear LHM and nonlinear RHM was presented
in~\cite{Shadr4,Darmanyan2}.

\subsection{Nonlinear planar waveguide}

In this section we  consider a  sandwich type structure, which
consists of a dielectric layer with  left-handed properties
surrounded by  Kerr-like nonlinear dielectrics~\cite{Shadr3}. The
presence of two interfaces in this configuration (at $x=0$ and
$x=h$) could lead to  two types of guided waves. The first type is
represented by an internal wave localized in the layer.  We will
refer to these types of solutions as non-linear guided modes (NLGM).
The second type is made up of two  coupled surface waves and they
will be referred to as non-linear surface waves (NLSW).

\subsubsection{Dispersion relations of nonlinear surface TE-waves}

The equation (\ref{eq815})  describing the  transverse electric
field profile of the wave in a three-layer structure can be
decomposed into a system of simple equations. If the condition $\mu
_{i}\varepsilon _{nl}^{(i)}>0$ holds, then the solution of these
equations is
\begin{equation*}
\Phi (x)=\left\{
\begin{array}{lr}
\alpha _{1}^{-1/2}\sec \mathrm{h}[p_{1}(x-x_{1})], & x\,<0 \\
A\exp (-\kappa x)+B\exp (\kappa x), & 0<x<h \\
\alpha _{2}^{-1/2}\sec \mathrm{h}[p_{2}(x-x_{2})], & x>h
\end{array}
\right. ,
\end{equation*}
where $p_{1}^{2}=\beta ^{2}-k_{0}^{2}\varepsilon _{1}\mu _{1}$, $
p_{2}^{2}=\beta ^{2}-k_{0}^{2}\varepsilon _{2}\mu _{2}$, $i=1,2$ and
$\kappa ^{2}=\beta ^{2}-k_{0}^{2}\varepsilon _{s}\mu _{s}>0$.  The
signs of $p_{1}^{2}$ and $p_{2}^{2}$ are chosen to be positive to
guarantee validity of the boundary conditions~(\ref{eqA13}). The
sign of $\kappa ^{2}$ is also chosen to be positive
(see~(\ref{eq815})), as we will see below such a choice corresponds
to  the case of surface waves. For $\mu _{i}\varepsilon
_{nl}^{(i)}<0$\ we have
\begin{equation*}
\Phi (x)=\left\{
\begin{array}{lr}
\pm |\alpha _{1}|^{-1/2}\mathrm{co}\sec \mathrm{h}[p_{1}(x-x_{1})],
& x\,<0
\\
A\exp (-\kappa x)+B\exp (\kappa x), & 0<x<h \\
\pm |\alpha _{2}|^{-1/2}\mathrm{co}\sec \mathrm{h}[p_{2}(x-x_{2})],
& x>h
\end{array}
\right. .
\end{equation*}
There $x_{1,2}$ are the constants of integration corresponding to
the co-ordinates of the maximum of the transverse profile of the
electric field. The continuity conditions  of the electric and
magnetic field components at the interfaces~( \ref{eqA12}) result in
a homogeneous system of linear equations. There are nontrivial
solutions of this system if the determinant is zero, which leads to
the following dispersion relation:
\begin{equation}
e^{2\kappa h}\left( 1+\mu _{s}\tilde{p}_{1}/\kappa \right) \left(
1+\mu _{s} \tilde{p}_{2}/\kappa \right) =\left( 1-\mu
_{s}\tilde{p}_{1}/\kappa \right) \left( 1-\mu
_{s}\tilde{p}_{2}/\kappa \right),  \label{eqB29}
\end{equation}
where $\tilde{p}_{1,2}=\mu _{1,2}^{-1}p_{1,2}\tanh
(p_{1,2}x_{1,2})$. Introducing new variables  $\phi _{1}$ and $\phi
_{2}$\ as $\tanh (\phi _{1}/2)=\mu _{s}\tilde{p}_{1}/\kappa $,
$\tanh (\phi _{2}/2)=\mu _{s}\tilde{p} _{2}/\kappa $, we represent
relation~(\ref{eqB29}) in the form:
\begin{equation}
2\kappa h+\phi _{1}+\phi _{2}=0.  \label{eqV210}
\end{equation}
Since $\kappa h$ is positive, at least one of $\phi _{1}$ and $\phi
_{2}$ must be negative.  We define  the amplitudes of nonlinear
surface waves  at the corresponding  interfaces as:  $ A_{s1}=\alpha
_{1}^{-1/2}\sec \mathrm{h}(p_{1}x_{1}),~A_{s2}=\alpha
_{2}^{-1/2}\sec \mathrm{h}[p_{2}(h-x_{2})]$. The connection between
parameters $x_{1}$ and $x_{2}$ can be found from the condition
describing coupling of these two surface waves:
\begin{equation}
A_{s2}=A_{s1}\left[ \cosh \kappa h-(\mu _{s}\tilde{p}_{2}/\kappa
)\sinh \kappa h\right].  \label{eqD211}
\end{equation}
Taking into account the relation~(\ref{eqD211}) we conclude that the
dispersion relation~(\ref{eqB29}) contains $x_{1}$ as a free
parameter.  On the other hand, parameter $A_{s1}$ is defined in
terms of $x_{1}$, therefore the value of the electric field   at
$x=0$ can be chosen as a free parameter instead of $ x_{1}$.
Finally, equation~(\ref{eqB29}) can be considered as an implicit one
parameter expression for describing the dependance of   the
propagation constant on the frequency $\beta =\beta (\omega
;A_{s1})$.

\subsubsection{Dispersion relations of nonlinear guided waves}

Analysis of the nonlinear guided wave  is based on equation
(\ref{eq815}). In contrast to the previous  case,  here  we choose
the sign of the coefficient in front of the function $\Phi$ in the
equation~(\ref{eq815}) to be negative  i.e. $k_{0}^{2}\varepsilon
_{s}\mu _{s}-\beta ^{2}=\kappa_{1} ^{2}>0$. If $\mu _{i}\varepsilon
_{nl}^{(i)}>0$, the solution of equation (\ref{eq815}) is
represented by the expression
\begin{equation*}
\Phi (x)=\left\{
\begin{array}{lr}
\alpha _{1}^{-1/2}\sec \mathrm{h}[p_{1}(x-x_{1})], & x<0 \\
A\exp (i\kappa_{1} x)+B\exp (-i\kappa_{1} x), & 0<x<h \\
\alpha _{2}^{-1/2}\sec \mathrm{h}[p_{2}(x-x_{2})], & x>h
\end{array}
\right. .
\end{equation*}
In this case the dispersion relation for NLGW reads as
\begin{equation}
e^{2i\kappa_{1} h}\frac{\left( 1-i\mu _{s}\tilde{p}_{1}/\kappa_{1}
\right) \left( 1-i\mu _{s}\tilde{p}_{2}/\kappa_{1} \right) }{\left(
1+i\mu _{s}\tilde{p} _{1}/\kappa_{1} \right) \left( 1+i\mu
_{s}\tilde{p}_{2}/\kappa_{1} \right) }=1, \label{eqB214}
\end{equation}
or
\begin{equation}
\tan (\kappa_{1} h)=\frac{\mu _{s}\kappa_{1} \left(
\tilde{p}_{1}+\tilde{p} _{2}\right) }{\kappa_{1} ^{2}-\mu
_{s}^{2}\tilde{p}_{1}\tilde{p}_{2}}. \label{eqB215}
\end{equation}
If we define the phases $\phi _{1}$ and $\phi _{2}$ by the formulae:
$\tan (\phi _{1}/2)=\mu _{s}\tilde{p}_{1}/\kappa_{1} $, $\tan (\phi
_{2}/2)=\mu _{s} \tilde{p}_{2}/\kappa_{1} $, then equation
(\ref{eq814}) can be written as
\begin{equation}
2\kappa_{1} h=\phi _{1}+\phi _{2}+2\pi m,~m=0,1,2,... \label{eqB217}
\end{equation}
As in to the case of linear waveguides, this expression shows that
the full phase shift of the zigzag wave~\cite{Integ} consists of
contributions  from the linear medium of the LHM dielectric slab and
the phase shifts $\ -\phi _{1}$ and $-\phi _{2}$, which occur in the
total internal reflection at the linear-non-linear interfaces. This
nonlinear phase shift can be named the \emph{non-linear Goos-Hanchen
effect}~\cite{GH1,GH2,R43}

Analysis of  TM waves is presented  in~\cite{Darman,Aguano}. The
corresponding dispersion relation and   NLGW modes were studied
in~\cite{Shadr4}. In particular, symmetric, asymmetric, and
antisymmetric, forward and backward modes have been found.

\subsection{Wave refraction and reflection at thin-film on interface
between two dielectrics}

Let us consider the case where a thin film is inserted between two
isotropic media. The optical properties of these media are described
by the permittivity and permeability: $\varepsilon _{1}$ and $\mu
_{1}$ at $x<0$ , $\varepsilon _{2}$ and $\mu _{2}$ at $x>0$. Let us
assume that the thin film of metamaterial separates these two media.
The optical properties of the metamaterial are described by the
polarization $\mathbf{P}^{(s)}$ and the magnetization
$\mathbf{M}^{(s)}$. We assume that the width of film $l_{f}$ is much
less than the carrier wave length of a quasiharmonic electromagnetic
wave. To describe the propagation of the electromagnetic wave
through thin film  it is sufficient to take account of the jump in
the electric and magnetic field components induced by polarization
and magnetization of the film. Following to~\cite{Rupaso21,Rupaso22}
we can approximate transverse distribution of polarization and
magnetization in the film by a delta-function located at $ x=0$.
This approximation leads to the following jump conditions
\begin{equation}
E_{y}(0-)-E_{y}(0+)=\frac{4\pi }{c}\frac{\partial }{\partial t}
M_{z}^{(s)},\quad H_{z}(0-)-H_{z}(0+)=\frac{4\pi }{c}\frac{\partial
}{
\partial t}P_{z}^{(s)},  \label{BC1}
\end{equation}
for the TE wave, and
\begin{subequations}
\begin{equation}
H_{y}(0-)-H_{y}(0+)=-\frac{4\pi }{c}\frac{\partial }{\partial t}
P_{z}^{(s)},\quad E_{z}(0-)-E_{z}(0+)=-\frac{4\pi }{c}\frac{\partial
}{
\partial t}M_{z}^{(s)},  \label{BC2}
\end{equation}
for the TM wave. The expressions~(\ref{BC1}) and~(\ref{BC2}) are
modifications the corresponding boundary relations used
in~\cite{Rupaso21,Rupaso22,Bash88,Van,BBM}. These new relations are
applicable both for    the case of continuum waves and for
ultra-short or extremely short pulses.

It follows from~(\ref{BC1}) and~(\ref{BC2}) that the electric and
magnetic field components have different values in the media
surrounding the film. The values of $E$ and $H$ (i.e., field values
acting on nanostructures  (meta-atoms) of the film), are naturally
defined as follows:
\end{subequations}
\begin{equation}
E_{a}(x=0)=0.5\left[ E_{a}(0-)+E_{a}(0-)\right]  \label{eq6D1}
\end{equation}
\begin{equation}
H_{a}(x=0)=0.5\left[ H_{a}(0-)+H_{a}(0-)\right]  \label{eq6D2}
\end{equation}
where $a=y$ or $z$.  In the case of a nonmagnetic film,
using~(\ref{BC1}), we obtain the well known
result~\cite{Rupaso21,Rupaso22}  for a TE-wave
$E_{y}(x=0)=E_{y}(0-)=E_{y}(0-).$

Let us consider a plane wave which has  normal incidence to the
interface.  For the sake of simplicity we assume that the two
dielectric media surrounding  the film are dispersionless. Then in
the general case the electric field
 can be represented in terms of incident, reflected
and transmitted waves:
\begin{equation}
E(x,t)=\left\{
\begin{array}{cc}
E_{in}(t-x/V_{1})+E_{ref}(t+x/V_{1}) & x<0 \\
E_{tr}(t-x/V_{2}) & x>0
\end{array}
\right. .  \label{eq6}
\end{equation}
Here $V_{1}$\ is group velocity in the dielectric at $x<0$, and
$V_{2}$ is group velocity in dielectric at $x>0$. Using  the jump
conditions ~(\ref {BC1}), (\ref{BC2}) and equation
$E_{y,x}=-c^{-1}H_{z,t}$ we find the analog of the Fresnel
relations:
\begin{eqnarray}
E_{tr}(t) &=&\frac{2V_{2}}{V_{1}+V_{2}}E_{in}-\frac{4\pi V_{2}}{
c(V_{1}+V_{2})}\left\{ \frac{\partial }{\partial
t}M_{z}^{(s)}+\frac{V_{1}}{c
}\frac{\partial }{\partial t}P_{y}^{(s)}\right\} ,  \label{eq10} \\
E_{ref}(t) &=&\frac{V_{2}-V_{1}}{V_{1}+V_{2}}E_{in}+\frac{4\pi
V_{1}}{ c(V_{1}+V_{2})}\left\{ \frac{\partial }{\partial
t}M_{z}^{(s)}-\frac{V_{2}}{c }\frac{\partial }{\partial
t}P_{y}^{(s)}\right\} ,  \label{eq11}
\end{eqnarray}
where $P_{y}^{(s)}=P_{y}(t,x=0)$ and $M_{z}^{(s)}=M_{z}(t,x=0)$ are
surface polarization and magnetization of the thin film
respectively.

For the TM-wave, using a similar approach, we can obtain following
the Fresnel relations:
\begin{eqnarray}
H_{tr}(t) &=&\frac{2\varepsilon _{2}V_{2}}{\varepsilon
_{1}V_{1}+\varepsilon _{2}V_{2}}H_{in}-\frac{4\pi \varepsilon
_{1}\varepsilon _{2}V_{1}V_{2}}{ c(\varepsilon _{1}V_{1}+\varepsilon
_{2}V_{2})}\left\{ \frac{\partial }{
\partial t}M_{y}^{(s)}-\frac{c}{\varepsilon _{1}V_{1}}\frac{\partial }{
\partial t}P_{z}^{(s)}\right\} ,  \label{e12a} \\
H_{ref}(t) &=&\frac{\varepsilon _{2}V_{2}-\varepsilon
_{1}V_{1}}{\varepsilon _{1}V_{1}+\varepsilon
_{2}V_{2}}H_{in}+\frac{4\pi \varepsilon _{1}\varepsilon
_{2}V_{1}V_{2}}{c(\varepsilon _{1}V_{1}+\varepsilon
_{2}V_{2})}\left\{ \frac{\partial }{\partial t}M_{y}^{(s)}+\frac{c}{
\varepsilon _{1}V_{1}}\frac{\partial }{\partial
t}P_{z}^{(s)}\right\} , \label{e12b}
\end{eqnarray}
where the magnetic field amplitudes are defined by:
\begin{equation*}
H_{y}(x,t)=\left\{
\begin{array}{cc}
H_{in}(t-x/V_{1})+H_{ref}(t+x/V_{1}), & x<0 \\
H_{tr}(t-x/V_{2}), & x>0
\end{array}
\right. .
\end{equation*}
The electric field  can be  defined from the equation
$H_{y,x}=c^{-1}E_{z,t}$.

In what follows we will consider the case of a TE wave since there
is no differences between TE and TM waves in case of normal incident
waves. To obtain the polarization and magnetization of the thin film
it is necessary to choose an appropriate model for the meta-atoms of
the film. We start from the hypotheses that the linear response of
film is determined by the following expressions: $ \varepsilon
_{f}(\omega )=1-(\omega _{p}/\omega )^{2}$, $\mu _{f}(\omega
)=1-\beta _{m}\omega ^{2}/(\omega _{T}^{2}-\omega ^{2})$. This
response follows from a standard linear Lorenz model which considers
meta-atoms as linear oscillators forced by electric and magnetic
fields. One natural generalization of such a model is to take into
account anharmonicity of the electrical oscillations in
nanostructures. This leads to the simplest model for nonlinear
response of metamaterials:
\begin{equation}
P_{,tt}+\omega _{d}^{2}P+\Gamma _{e}P_{,t}+g_{p}P^{3}=(\omega
_{p}^{2}/4\pi )E,  \label{eq141}
\end{equation}
\begin{equation}
M_{,tt}+\omega _{T}^{2}P+\Gamma _{m}M_{,t}=-(\beta _{m}/4\pi
)H_{,tt}, \label{eq142}
\end{equation}
where $\omega _{d}$ is frequency of the dimensional quantization due
to confinement of the plasma in the nanostructures. Losses in the
metallic nanostructures are taken into account by the parameters
$\Gamma _{e}$\ and $\Gamma _{m}$. Let us  normalize  the variables
in equations~(\ref{eq10}, \ref{eq11}) and (\ref{eq141}, \ref{eq142})
\begin{eqnarray*}
e_{tr} &=&E_{tr}/4\pi P_{0},e_{ref}=E_{ref}/4\pi
P_{0},e_{in}=E_{in}/4\pi
P_{0}, \\
q &=&P/P_{0},m=M/\beta _{m}\sqrt{\varepsilon _{2}}P_{0},\tau =\omega
_{p}t.
\end{eqnarray*}
The system of the normalized equations takes the following form
\begin{equation}
e_{tr}(\tau)=\frac{2\sqrt{\varepsilon _{2}}}{\sqrt{\varepsilon
_{1}}+\sqrt{ \varepsilon _{2}}}e_{in}(\tau)-\frac{l_{f}\omega
_{p}}{c(\sqrt{\varepsilon _{1}} +\sqrt{\varepsilon _{2}})}\left\{
\frac{\partial q}{\partial \tau}+\beta _{m} \sqrt{\varepsilon
_{1}\varepsilon _{2}}\frac{\partial m}{\partial \tau}\right\} ,
\label{eqD1}
\end{equation}
\begin{equation}
e_{ref}(\tau)=\frac{\sqrt{\varepsilon _{2}}-\sqrt{\varepsilon
_{1}}}{\sqrt{ \varepsilon _{1}}+\sqrt{\varepsilon
_{2}}}e_{in}(\tau)-\frac{l_{f}\omega _{p}}{ c(\sqrt{\varepsilon
_{1}}+\sqrt{\varepsilon _{2}})}\left\{ \frac{\partial q}{
\partial \tau}-\beta _{m}\sqrt{\varepsilon _{1}\varepsilon _{2}}\frac{\partial m
}{\partial \tau}\right\} ,  \label{eqD2}
\end{equation}
\begin{equation}
q_{,\tau \tau }+(\omega _{d}/\omega _{p})^{2}q+(\Gamma _{e}/\omega
_{p})q_{,\tau }+(g_{p}P_{0}^{2}/\omega
_{p}^{2})q^{3}=0.5(e_{in}(\tau)+e_{tr}(\tau)+e_{ref}(\tau)),
\label{eqD3}
\end{equation}
\begin{equation}
m_{,\tau \tau }+(\omega _{T}/\omega _{p})^{2}m+(\Gamma _{m}/\omega
_{p})m_{,\tau }=-0.5\left[ e_{tr}(\tau)+\sqrt{\varepsilon
_{1}/\varepsilon _{2}} \left( e_{in}(\tau)+e_{ref}(\tau)\right)
\right] _{,\tau \tau }.  \label{eqD4}
\end{equation}
Inhomogeneous broadening of the resonance line can be taken into
account if we replace $q$ and $m$ in (\ref{eqD1}- \ref{eqD4})  by
$\left\langle q\right\rangle $\ and $\left\langle m\right\rangle $.
Here corner brackets $\left\langle \cdot\cdot  \right\rangle$ denote
averaging over different frequencies $\omega _{d}$ and $\omega_{T}$.

We assume that the pulse duration is much shorter than the
characteristic plasmonic oscillation damping time, and neglect the
dissipation terms in (\ref{eqD1}- \ref{eqD4}). Furthermore, if the
slowly varying envelope approximation is acceptable, we can reduce
the system of these equations and consider the following system
\begin{eqnarray}
\mathcal{E}_{tr}(t) &=&\frac{2q_{1}(\omega _{0})}{q_{1}(\omega
_{0})+q_{2}(\omega _{0})}\mathcal{E}_{in}+\frac{4\pi
ik_{0}n_{at}l_{f}}{ q_{1}(\omega _{0})+q_{2}(\omega _{0})}\left[
k_{0}\left\langle \mathcal{P} \right\rangle +q_{1}(\omega
_{0})\left\langle \mathcal{M}\right\rangle
\right] ,  \label{eqD81} \\
\mathcal{E}_{ref}(t) &=&\frac{q_{2}(\omega _{0})-q_{1}(\omega
_{0})}{ q_{1}(\omega _{0})+q_{2}(\omega
_{0})}\mathcal{E}_{in}+\frac{4\pi ik_{0}n_{at}l_{f}}{q_{1}(\omega
_{0})+q_{2}(\omega _{0})}\left[ k_{0}\left\langle
\mathcal{P}\right\rangle -q_{2}(\omega _{0})\left\langle
\mathcal{M}\right\rangle \right]  \label{eqD82}
\end{eqnarray}
\begin{equation}
\mathcal{P}_{,t}-i(\omega _{0}-\omega
_{d})\mathcal{P-}(3ig_{p}/2\omega
_{0})|\mathcal{P}|^{2}\mathcal{P}=i(\omega _{p}^{2}/8\pi \omega
_{0}) \mathcal{E}_{f},  \label{eqD83}
\end{equation}
\begin{equation}
\mathcal{M}_{,t}-i(\omega _{0}-\omega _{T})\mathcal{M}=i(\beta
_{m}\omega _{0}/8\pi )\mathcal{H}_{f},  \label{eqD84}
\end{equation}
Here $k_{0}=\omega _{0}/c$, $\omega _{0}$ is the carrier wave
frequency, $ \mathcal{E}_{in}$, $\mathcal{E}_{tr}$ and
$\mathcal{E}_{ref}$ are envelopes of the incident, transmitted and
reflected pulses, $\mathcal{P}$ and $ \mathcal{M}$\ are envelopes of
the polarization and magnetization. The wavenumbers $q_{1,2}$ are
determined as $q_{1,2}=k_{0}\sqrt{\varepsilon _{1,2}}\cos \theta
_{1,2}$; $\theta _{1}$ and $\theta _{2}$ are  incident and
refraction angles respectively;   $n_{at}$ is volume density of
meta-atoms. The envelopes of electric $\mathcal{E}_{f}$ and magnetic
$\mathcal{H}_{f}$ fields, acting on nanostructures (meta-atoms) of
the thin film can be represented as follows:
\begin{equation*}
\mathcal{E}_{f}=\left[
\mathcal{E}_{tr}+\mathcal{E}_{in}+\mathcal{E}_{ref} \right] /2,\quad
\mathcal{H}_{f}=\left[ \sqrt{\varepsilon _{2}}\mathcal{E}
_{tr}+\sqrt{\varepsilon
_{1}}(\mathcal{E}_{in}-\mathcal{E}_{ref})\right] /2.
\end{equation*}

Using the normalized variables $e_{tr}=\mathcal{E}_{tr}/A_{0}$,
$e_{ref}= \mathcal{E}_{ref}/A_{0},e_{in}=\mathcal{E}_{in}/A_{0}$,
$q=\mathcal{P} /P_{0},m=\mathcal{M}/M_{0},\tau =\omega _{p}t$, and
introducing parameters  $P_{0}=(\omega _{p}/8\pi \omega _{0})A_{0}$
and  $M_{0}=(\beta _{m}\omega _{0}\sqrt{\varepsilon _{2}}/8\pi
\omega _{p})A_{0}$, we present the dimensionless system of equations
describing the interaction of the electromagnetic pulse with a thin
film of metamaterial
\begin{eqnarray}
e_{tr}(t)
&=&\frac{2n_{1}}{n_{1}+n_{2}}e_{in}+\frac{ik_{0}n_{at}l_{f}}{
2(n_{1}+n_{2})}\left[ \frac{\omega _{p}}{\omega _{0}}\left\langle
q\right\rangle +\beta _{m}n_{1}n_{2}\frac{\omega _{0}}{\omega _{p}}
\left\langle m\right\rangle \right] ,  \label{eqD91} \\
e_{ref}(t)
&=&\frac{n_{2}-n_{1}}{n_{1}+n_{2}}e_{in}+\frac{ik_{0}n_{at}l_{f}}{
2(n_{1}+n_{2})}\left[ \frac{\omega _{p}}{\omega _{0}}\left\langle
q\right\rangle -\beta _{m}n_{2}n_{2}\frac{\omega _{0}}{\omega _{p}}
\left\langle m\right\rangle \right]  \label{eqD92}
\end{eqnarray}
\begin{eqnarray}
q_{,\tau } &=&i\omega _{p}^{-1}(\omega _{0}-\omega
_{d})q+ig_{2}|q|^{2}q+i\left( e_{tr}+e_{in}+e_{ref}\right) /2,
\label{eqD93}
\\
m_{,\tau } &=&i\omega _{p}^{-1}(\omega _{0}-\omega _{T})m+i\left[
e_{tr}+(n_{1}/n_{2})(e_{in}-e_{ref})\right] /2.  \label{eqD94}
\end{eqnarray}
Here $n_{1.2}=\sqrt{\varepsilon _{1,2}}$ are refractive indices of
the dielectric media. The constant of anharmonicity $g_{2}$, used in
these equations, is defined as $ g_{2}=(3g_{p}/2\omega _{0}\omega
_{p})(\omega _{p}/8\pi \omega _{0})^{2}A_{0}^{2}$. The free
parameter $A_{0}$ can be  represented in terms of the peak intensity
of the incident pulse.

In the case of  a thin film with no magnetization  the jump
conditions ~(\ref{BC1}) results in the relation
$e_{tr}=e_{in}+e_{ref}$. Equations~(\ref{eqD91})-(\ref{eqD94})in
this case transform into  the system of equations describing the
refraction of the ultra-short electromagnetic pulse on the thin film
with embedded nanoparticles, which was considered in~\cite{GILMSS}.

\subsection{Optical bistability in nonlinear layered structures}

A linear layer of dielectric material acts as a resonator for an
incident light beam. The transmissivity of this layer is determined
by the thickness and refractive index of the layer material and by
the wavelength of the incident light. If such a dielectric is
nonlinear then its optical thickness depends on the intensity of the
light field. Therefore, transmissivity of this layer depends on the
light intensity. Let us consider the resonance reflection condition
in the linear limit. Gradual increase of light intensity will be
followed by a drastic increase in the transmissivity if the light
intensity reaches a certain threshold level. In other words, the
system switches from an opaque to a transparent state. This is the
phenomena of optical bistability~\cite{Gibbs}. Phenomena of
bistability in thin films were studied in~\cite{Bash88,Ben}. The
more realistic case of a film with nonzero thickness was considered
in~\cite{Koep,Peshle}. The switching condition is determined by the
phase difference of the waves reflected from the front and back
boundaries of the layer. Let us consider a sandwich structure with a
metamaterial film and a layer of nonlinear dielectric. The presence
of the metamaterial film results in a change in the effective
thickness of the resonator. If the refractive index of such a film
is negative, then the effective optical thickness of the
sandwich-resonator is less than the optical thickness of the
resonator without film. Therefore,  the switching condition is
realized at a higher level of light intensity~\cite{Litch}.

It was also shown in~\cite{Litch} that switching light intensity
depends on the incident beam direction. Switching intensity is
higher for the case where the light is entering the structure from
the side covered by the metamaterial film than for the case of
opposite beam direction.

The hysteresis loop is an important characteristic of bistable
devices. Switching thresholds from opaque to transparent state and
from transparent to opaque state are different. The hysteresis width
is
 the difference of these two switching thresholds.
The presence of a metamaterial film, as  was shown in~\cite{Litch},
results in an increase of the hysteresis width.

\section{Parametric interactions}

Transformation of frequency of an electromagnetic wave propagating
in a non-linear medium is one of the most fundamental
effects~\cite{Shen} forming a basis for a broad range of nonlinear
optics phenomena. Three waves interaction is the simplest
representative of such a class of phenomena.  When  frequencies of
two waves from an  interacting wave triad are equal, the frequency
of the third wave is two times higher.  This is known as  second
harmonic generation (SHG). Stimulated Raman scattering is another
example of three-wave interaction. In this case one of the waves is
of acoustic nature  and the incident and scattered waves are
electromagnetic. Four wave interaction is a more complicated process
of parametric interactions. Third harmonic generation (THG) is a
typical representative of these processes.

\subsection{Three-wave parametric interaction}

Let as assume that the non-linear characteristics of a medium to be
described by non-linear susceptibility of second order $\chi
^{(2)}$. Such a medium is commonly known  as a quadratic non-linear
medium. We consider  waves with carrier frequencies $\omega _{1}$
and $\omega _{2}$ propagating along the $z$ axis. The polarization
of such a medium is  a quadratic function of the electrical field,
therefore waves with frequencies  $\omega _{s}=\omega _{1}\pm \omega
_{2}$, $2\omega _{1}$, and $ 2\omega _{2}$ must be generated in such
a medium. These waves, when their amplitudes increase sufficiently
to be involved in the process of nonlinear interaction, can generate
new waves with the frequencies $2\omega _{1}\pm \omega _{2}$, $
\omega _{1}\pm 2\omega _{2}$, $4\omega _{1}$, and $4\omega _{2}$ so
on. However in a dispersive medium all these processes are not
equally efficient. There is an important \emph{phase matching
condition}, which selects and emphases a certain type of interaction
of three waves, leaving all others unaffected. Sometimes such a
phase matching occurs for  waves propagating in the same direction -
the case of collinear parametric interaction. In this case the
distance at which the interaction of waves is taking place can be
made sufficiently long and, consequently, the effective frequency
transformation will take place.

The equations describing  three wave interaction ($\omega
_{3}=\omega _{1}+\omega _{2}$) in a $\chi ^{(2)}$-medium in the
slowly varying envelope and phase approximation, taking into account
the group-velocity dispersion, can be written in following form
\cite{Shen}:
\begin{eqnarray}
\left( \hat{k}_{1}\frac{\partial }{\partial
z}+\frac{1}{v_{1}}\frac{\partial }{\partial t}\right)
\mathcal{E}_{1}-\frac{D_{1}}{2}\frac{\partial ^{2}
\mathcal{E}_{1}}{\partial t^{2}} &=&i\frac{2\pi \omega _{1}^{2}\mu
(\omega
_{1})}{c^{2}k_{1}}\tilde{P}_{NL}(\omega _{1})\exp (-ik_{1}z) \\
\left( \hat{k}_{2}\frac{\partial }{\partial
z}+\frac{1}{v_{2}}\frac{\partial }{\partial t}\right)
\mathcal{E}_{2}-\frac{D_{2}}{2}\frac{\partial ^{2}
\mathcal{E}_{2}}{\partial t^{2}} &=&i\frac{2\pi \omega _{2}^{2}\mu
(\omega
_{2})}{c^{2}k_{2}}\tilde{P}_{NL}(\omega _{2})\exp (-ik_{2}z)  \label{eq1} \\
\left( \hat{k}_{3}\frac{\partial }{\partial
z}+\frac{1}{v_{3}}\frac{\partial }{\partial t}\right)
\mathcal{E}_{3}-\frac{D_{3}}{2}\frac{\partial ^{2}
\mathcal{E}_{3}}{\partial t^{2}} &=&i\frac{2\pi \omega _{3}^{2}\mu
(\omega _{3})}{c^{2}k_{3}}\tilde{P}_{NL}(\omega _{3})\exp
(-ik_{3}z).
\end{eqnarray}
Here $k_{j}^{2}$ is defined as $k_{j}^{2}=(\omega
_{j}/c)^{2}\varepsilon (\omega _{j})\mu (\omega _{j})$;
$\hat{k}_{j}$ is sign of square root of $k_{j}^{2}$; and
\begin{eqnarray*}
\tilde{P}_{NL}(\omega _{1}) &=&\chi ^{(2)}(\omega _{1};\omega
_{3},-\omega
_{2})\mathcal{E}_{3}\mathcal{E}_{2}^{\ast }\exp [iz(k_{3}-k_{2})], \\
\tilde{P}_{NL}(\omega _{2}) &=&\chi ^{(2)}(\omega _{2};\omega
_{3},-\omega
_{1})\mathcal{E}_{3}\mathcal{E}_{1}^{\ast }\exp [iz(k_{3}-k_{1})], \\
\tilde{P}_{NL}(\omega _{3}) &=&\chi ^{(2)}(\omega _{3};\omega
_{1},\omega _{2})\mathcal{E}_{1}\mathcal{E}_{2}\exp
[iz(k_{1}+k_{2})].
\end{eqnarray*}
If $\omega _{1}=\omega _{2}$, $\omega _{3}=2\omega _{1}$ we have
second harmonic generation. If  $\omega _{1}=\omega _{s}$, $\omega
_{2}=\omega _{i}$, and $\omega _{3}=\omega _{p}$ ($\omega
_{p}=\omega _{s}+\omega _{i}$), than we are dealing with parametric
amplification phenomena. The signal wave  is amplified by taking
energy from the pump wave with the help of the idler wave, playing
the role of a mediator in this energy transfer. The idler wave here
corresponds to $\mathcal{E}_{2}$ and  the pump wave corresponds to $
\mathcal{E}_{3}$.

The zero phase mismatch condition $\Delta k=k_{p}-k_{s}-k_{i}$ means
that the vector of the pump wave $\mathbf{k}_{p}$ is equal to the
vector $\mathbf{k}_{s}+ \mathbf{k}_{i}$. The configuration of these
vectors defines the energy flow directions. If the nonlinear medium
is characterized by a positive refraction index, then the vector
$\mathbf{k}_{p}$ and the Poynting vectors associated with the
interacting waves have the same orientation. Opposite directionality
takes place if the nonlinear medium is characterized by negative
refraction index~\cite{Agran,Kivsh05,Popov,Popov1}. Currently the
negative refractive property in the optical domain has been realized
by use of simultaneous resonance for electric and magnetic field
components in metallic nanostructures embedded to a host medium. The
resonance frequencies in these materials are close to each other and
they divide the frequency range into two domains. The medium
responds to external waves as a negative refractive index material
if the wave carrier frequency is below both resonance frequencies.
The medium responds to external waves as a positive refractive index
material if the wave carrier frequency is above both resonance
frequencies. In  the case of a three wave interaction, wave vectors
corresponding to the  waves with frequencies from the frequency
region of negative refraction index are directed according to the
zero phase mismatch condition, but the associated Poynting vectors
are have opposite orientation.

Let us consider parametric amplification. It is convenient to
introduce  normalized variables $\mathcal{E}_{1}=A_{0}\gamma
_{1}e_{s}$ , $\mathcal{E} _{2}=A_{0}\gamma _{2}e_{2}$,
$\mathcal{E}_{3}=A_{0}\gamma _{2}e_{1}$,where $ \gamma _{j}=2\pi
\omega _{1}\chi ^{(2)}(\omega _{1})c^{-1}\sqrt{\mu (\omega
_{j})/\varepsilon (\omega _{j})}$. The system of equations describing the parametric
amplification can be  rewritten as follows:
\begin{eqnarray}
\left( \frac{\partial }{\partial z}+\frac{1}{v_{1}}\frac{\partial
}{\partial t}\right) e_{1}+i\frac{D_{1}}{2}\frac{\partial
^{2}e_{1}}{\partial t^{2}}
&=&ige_{2}e_{s}\exp (-i\Delta kz) \\
\left( \frac{\partial }{\partial z}+\frac{1}{v_{2}}\frac{\partial
}{\partial t}\right) e_{2}+i\frac{D_{1}}{2}\frac{\partial
^{2}e_{2}}{\partial t^{2}}
&=&ige_{1}e_{s}^{\ast }\exp (+i\Delta kz)  \label{eq4} \\
\left( -\frac{\partial }{\partial z}+\frac{1}{vs}\frac{\partial
}{\partial t} \right) e_{s}+i\frac{D_{1}}{2}\frac{\partial
^{2}e_{s}}{\partial t^{2}} &=&ige_{1}e_{2}^{\ast }\exp (+i\Delta
kz).
\end{eqnarray}
Here $\Delta k$ is defined as $\Delta k=k_{p}-k_{s}-k_{i}$\ and
$g=(\gamma _{1}\gamma _{2}\gamma _{3})^{1/2}A_{0}$\ is a coupling
constant. The value $A_{0}$ is introduced for appropriate
normalization of  electric fields of interacting waves.

In the general case the phase matching condition requires pump and
signal waves to  propagate in opposite directions. However, solitary
wave solutions of~(\ref{eq4}) are possible   if the power of the
interacting waves is sufficiently high. Such solitary wave solutions
can be considered as a bounded state of signal, idler and pump waves
propagating in the same direction. This bound state can be
interpreted as a wave trapping phenomena, where the   group velocity
of the signal wave changes sign due to a nonlinearly induced change
in the  value of the effective refractive index. Note that if the
phase mismatch is zero $\Delta k=0$ and the effect of group-velocity
dispersion is negligible (short sample), then equations~(\ref{eq4})
reduce to a system of equation integrable by use of the inverse
scattering transform~\cite{ZaMan,Kaup76,AbloSeg}. Therefore,
results previously obtained for the integrable case of three-wave
interaction can also be  used to analyze this special case of
three-wave interaction in LHM.

Negative refractive index materials based on plasmonic resonance in
metallic nanostructures have large losses. The feasibility of
parametric amplification as a means to compensate for dissipative
losses
 in LHM was considered in~\cite{PopSha3}. It was shown
that parametric amplifiers are capable of compensating for losses
and additionally that they can be  used as optical parametric
oscillators. The opposite directionality of  parametrically
interacting  waves in LHMs is an analog of distributed feedback. In
conventional materials distributed feedback is realized by adding
Bragg gratings.

The case of three-microwave interaction for  a strong pump wave and
two weak signals in metamaterials with nonlinear magnetic response
was theoretically studied in~\cite{Lapine}. In this case nonlinear
properties of the material were determined by the embedded LC
circuits with nonlinear resistance and capacitance.

\subsection{Second Harmonic Generation}

Second harmonic generation is a special  case of  three wave
interaction ($ \omega _{1}+\omega _{1}=\omega _{2}$) in a $\chi
^{(2)}$-medium. The equations of SHG in the case of  collinear
mismatch conditions $\Delta k=2k_{1}-k_{2}$\ follows from
(\ref{eq4}), if $e_{s}$  is chosen as $e_{s}=\tilde{e}_{1}$ for
fundamental  wave (with carrier frequency $\omega$) and if  $e_{1}$
in (\ref{eq4}) is replaced by $\tilde{e}_{2}$.  In this new notation
idler and pump waves correspond to a second harmonic:
\begin{eqnarray}
-\tilde{e}_{1,z}+v_{1}^{-1}\tilde{e}_{1,t}+i(D_{1}/2)\tilde{e}_{1,tt}
&=&ig
\tilde{e}_{2}\tilde{e}_{1}^{\ast }\exp (-i\Delta kz) \\
\tilde{e}_{2,z}+v_{2}^{-1}\tilde{e}_{2,t}+i(D_{2}/2)\tilde{e}_{2,tt}
&=&ig \tilde{e}_{1}^{2}\exp (+i\Delta kz)
\end{eqnarray}
Transformations  of variables $\tilde{e}_{1}=A_{10}e_{1}$,
$\tilde{e}_{2}=2A_{10}e_{2}\exp (+i\Delta kz)$  allow us to
represent these equations in the following standard form
\begin{eqnarray}
ie_{1,\zeta }+(\sigma /2)e_{1,\tau \tau }-e_{2}e_{1}^{\ast }&=&0,
\label{A51} \\
ie_{2,\zeta }+i\delta e_{2,\tau }-(\beta /2)e_{2,\tau \tau }-\Delta
e_{2}+e_{1}^{2}/2&=&0.  \label{A52}
\end{eqnarray}
Here $A_{10}$, $\zeta$ and  $\tau$ are defined as
$A_{10}=(gL)^{-1}$, $\zeta =z/L$, $\tau =(t+z/v_{1})/t_{p}$.
Normalizing characteristic parameters here are chosen as follows: $
L=t_{p}^{2}/|D_{1}|$ is dispersion length; $\delta
=Lt_{p}^{-1}(v_{1}^{-1}+v_{2}^{-1})$ is normalized group velocity
mismatch; $\Delta =\Delta kL$ is normalized phase mismatch; $t_{p}$
is characteristic time, which is not fixed yet; and $\sigma =
\mathrm{sgn}D_{1}$, $\beta =D_{2}/|D_{1}|$. Parameter $\delta $
takes into account  the walk-off effect for pump and harmonic pulses
that is due to the difference of the group velocities' directions
for the interaction waves. It should be pointed out that  in
contrast to the case of positive refractive index medium, this
parameter can not be zero in LHM's.

The form of the equations~(\ref{A51},\ref{A52}) is similar to that
of the equations describing the  evolution  of quadratic solitons as
a bounded complex of fundamental and second harmonic solitary waves
presented in~\cite{Malom,Skryab}. The only difference is in the
relative  sign of  group velocities of the pump and second harmonic
waves.

Using the equations (\ref{A51},\ref{A52}) and taking into account
the boundary condition $|e_{1,2}|^{2}\rightarrow
|e_{10,20}|^{2},~\tau \rightarrow \pm \infty$, here $e_{10}$ and
$e_{20}$ are constants (or $|e_{1,2}|^{2}\rightarrow 0~\tau
\rightarrow \pm \infty$), we arrive to the Manley-Rowe relation:
\begin{equation*}
\int\limits_{-\infty }^{\infty }\left(
|e_{2}|^{2}-(1/2)|e_{1}|^{2}\right) d\tau =\mathrm{const.}
\end{equation*}
Note that the sign of the second term in the integrand is negative
in contrast to the situation for SHG in nonlinear positive index
materials. This form of the Manley-Rowe relation reflects the fact
that the Poynting vectors, i.e., energy fluxes, for the fundamental
and the second harmonic are antiparallel, while their wave vectors
are parallel.

\subsubsection{Continuous wave limit}

The system of equations (\ref{A51},\ref{A52}) for continuous waves
reduces to following equations~\cite{Agran,Popov,Popov1}:
\begin{equation}
ie_{1,\zeta }-e_{1}^{\ast }e_{2}=0,\quad ie_{2,\zeta }-\Delta
e_{2}+e_{1}^{2}/2=0,  \label{CW}
\end{equation}
The boundary conditions for  a nonlinear plate of a finite width $l$
are as follows:
\begin{equation}
\mathrm{At\quad }\zeta =0\quad |e_{1}|=a_{0},\quad \mathrm{and\quad
at\quad } \zeta =l\quad |e_{2}|=0.  \label{Boud}
\end{equation}
The Manley-Rowe relation in this case reads
\begin{equation*}
2|e_{2}|^{2}-|e_{1}|^{2}=2c_{0}^{2}=\mathrm{const.}
\end{equation*}

Solutions of these equations~(\ref{CW}) were found
in~\cite{Popov,Popov1}. Spatial distributions of the harmonic and
pump wave are represented by the following expressions:
\begin{equation}
|e_{1}(\zeta )|=c_{0}\sqrt{2}\sec \left[ c_{0}(l-\zeta )\right]
,\quad |e_{2}(\zeta )|=c_{0}\tan \left[ c_{0}(l-\zeta )\right] .
\label{CWSol}
\end{equation}
The constant $c_{0}$ is defined by the relation
$a_{0}=c_{0}\sqrt{2}\sec \left[ c_{0}\zeta _{0}\right] $ that
follows from boundary conditions (\ref{Boud}). In the case of
quadratic-nonlinear PRI medium the amplitude of the pump wave
decreases with distance, while the second harmonic wave amplitude
increases. The energy of the pump wave is transferred into the
second harmonic wave and the energy fluxes of both waves are
aligned. In a quadratic-nonlinear NRI medium the energy fluxes are
oriented  in opposite directions, thus  both pump and second
harmonic wave amplitudes decrease with distance. Second harmonic
generation taking into account dissipation  has been investigated by
numerical simulation in the~\cite{Popov}.

\subsubsection{Large-mismatch limit}

During the last few years there has been growing interest in  three
wave interaction or its special case of  second harmonic generation
when  $\left\vert \Delta \right\vert
>>1$, (see review~\cite {Skryab}). This regime is known as the
large-mismatch limit, the cascading limit, or the effective Kerr
limit. In this limit, equations for SHG~(\ref{A51},\ref {A52}) can
be transformed into a nonlinear Schr\H{o}dinger equation for the
fundamental wave, and the amplitude of the second harmonic wave is
proportional to the squared amplitude of fundamental wave. Generally
quadratic nonlinearity is an order of magnitude stronger than cubic
nonlinearity. Therefore the large-mismatch limit case is very useful
for study of nonlinear cubic phenomena using quadratic nonlinearity.
This large-mismatch limit can, in a similar way,   be considered in
the case of a quadratic-nonlinear NRI medium. From (\ref{A52}) we
evaluate $e_{2}\approx e_{1}^{2}/2\Delta $,  substituting $e_{2}$ in
(\ref{A51}) results in
\begin{equation}
ie_{1,\zeta }+(\sigma /2)e_{1,\tau \tau }-(1/2\Delta
)|e_{1}|^{2}e_{1}=0. \label{NLS3}
\end{equation}

This equation (\ref{NLS3}) has soliton solutions when $\sigma =-1$.
The type of these soliton solutions is controlled by sign of
$\Delta$:  these are bright solitons if $\Delta
>0 $ and they are of dark type if $\Delta <0 $. Note that for the frequency of the
fundamental (pump) wave the medium  acts as left handed. On the
other hand the governing equation for the pump wave in a
large-mismatch limit is identical to the conventual NLS equation
written for a right handed material. This is an indication that  the
soliton properties of LHM and RHM are similar in the large-mismatch
limit.

\subsubsection{Solitary wave  solutions}

The solitary wave solutions of~(\ref{A51},\ref{A52}), represented by
two-frequency pulses,  have been considered in~\cite{MGK}. To find
such solutions it is convenient to introduce real variables for the
interacting waves
\begin{equation*}
e_{1}=a\exp (i\varphi _{1}),\quad e_{2}=b\exp (i\varphi _{2}).
\end{equation*}
Substitution of these expressions into (\ref{A51},\ref{A52}) leads
to
\begin{eqnarray}
a_{,\zeta }+(\sigma /2)\left( 2a_{,\tau }\varphi _{1,\tau }+a\varphi
_{1,\tau \tau }\right) &=&ab\sin \Phi ,  \label{eqA71} \\
2b_{,\zeta }+2\delta b_{,\tau }-\beta \left( 2b_{,\tau }\varphi
_{2,\tau
}+b\varphi _{2,\tau \tau }\right) &=&a^{2}\sin \Phi ,  \label{eqA72} \\
a\varphi _{1,\zeta }-(\sigma /2)\left( a_{,\tau \tau }-a\varphi
_{1,\tau
}\varphi _{1,\tau }\right) &=&-ab\cos \Phi ,  \label{eqA73} \\
2a\left( \varphi _{2,\zeta }+\delta \varphi _{2,\tau }\right) +\beta
\left( b_{,\tau \tau }+b\varphi _{2,\tau }\varphi _{2,\tau }\right)
+2\Delta b &=&a^{2}\cos \Phi ,  \label{eqA74}
\end{eqnarray}
where $\Phi =\varphi _{2}-2\varphi _{1}$. Assuming that phases  are
linear functions  of the following form $\varphi _{1}=K\zeta +\Omega
\tau $, $\varphi _{2}=2K\zeta +2\Omega \tau $ we obtain the mismatch
condition $\Phi =0$. The amplitude equations (\ref{eqA71}) and
(\ref{eqA72}) are equivalent if the frequency $\Omega $ is chosen to
be  $ \Omega =\delta /(\sigma +2\beta )$. Let us consider the
solitary wave regime represented by a bound state of two waves with
a fixed value of the amplitudes' ratio. These waves are defined by
functions of a single argument $a=a(\tau -\zeta /V)$, $b=b(\tau
-\zeta /V)$, with $V^{-1}=\sigma \Omega $ and  $b=fa$. Here $f$ is a
constant. Under such an assumption, the system  of equations
(\ref{eqA71}) and (\ref{eqA72}) is overdetermined (we have two
equations for one function $a=a(y)$, $y=\tau -\zeta /V$).   The
compatibility condition for these equation reads:
\begin{eqnarray*}
f^{2}&=&\sigma /2\beta >0,\\
K&=&\sigma \left( 3\beta \Omega ^{2}-4\delta \Omega -2\Delta \right)
/2(\beta +2\sigma ).
\end{eqnarray*}
and the equation for $a$ has the following form
\begin{equation*}
a_{,yy}-(2\sigma K+\Omega ^{2})a-2\sigma fa^{2}=0.
\end{equation*}
It follows from the compatibility condition that $\sigma$ and
$\beta$ must have the same sign.

Let us consider  the boundary condition $a\rightarrow 0$, $\partial
a/\partial y\rightarrow 0$ at $\tau \rightarrow \pm \infty $ , which
is consistent  with a solitary wave solution propagating on a zero
background (quadratic soliton -- see review~\cite {Skryab}). In the
case where  $\sigma =-1$ the solitary wave  solution of
(\ref{eqA71}-\ref{eqA74}) exists if $p=(\Omega ^{2}-2K)>0$. This
solution is a bright NRI medium soliton   with frequency components
of the following form
\begin{equation}
a(y)=\frac{(3p/4)\sqrt{2|\beta |}}{\cosh
^{2}[\sqrt{p}(y-y_{0})/2]},\quad b(y)=\frac{(3p/4)}{\cosh
^{2}[\sqrt{p}(y-y_{0})/2]}.  \label{eq22}
\end{equation}
The group velocity $V_s$ of the bright  soliton is fixed by system
parameters and defined as
\begin{equation*}
V_{s}^{-1}=\left( \sigma v_{2}^{-1}-2\beta v_{1}^{-1}\right)
/(\sigma +2\beta ).
\end{equation*}
Note, that in the case of a quadratic-nonlinear PRI medium, the
group velocity of a bright soliton is defined by the formula
$V_{s}^{-1}=\left( 2\beta v_{1}^{-1}-\sigma v_{2}^{-1}\right)
/(2\beta -\sigma )$. In the case where $\sigma =1$ the solitary wave
solution of (\ref{eqA71}-\ref{eqA74}) exists if $p=(\Omega
^{2}+2K)>0$.

If the background is nonzero, then the system of equations
(\ref{eqA71}-\ref{eqA74}) has dark soliton solutions. An example of
such a dark soliton at $\sigma =-1$ and $p=(\Omega ^{2}-2K)<0$ is
presented below:
\begin{eqnarray}
a(y) &=&(3|p|/4)\sqrt{2|\beta |}\left( 2/3\pm 2\sqrt{|p|}\sec
\mathrm{h}^{2}[
\sqrt{|p|}y/2]\right) ,  \label{A341} \\
b(y) &=&(3|p|/4)\left( 2/3\pm 2\sqrt{|p|}\sec
\mathrm{h}^{2}[\sqrt{|p|} y/2]\right).  \label{A342}
\end{eqnarray}
If the ratio of the coupled wave amplitudes is not fixed then the
system of equations (\ref{eqA71}-\ref{eqA74}) has double-hump
solitary waves, which are described by following expression
\begin{eqnarray}
a(y) &=&3p_{1}\sqrt{2\beta }\tanh \left( \sqrt{p_{1}}y\right) \sec
\mathrm{h}
\left( \sqrt{p_{1}}y\right) ,  \label{eq39a} \\
b(y) &=&3p_{1}\sec \mathrm{h}^{2}\left( \sqrt{p_{1}}y\right) .
\label{eq39b}
\end{eqnarray}
where $p_{1}=(\Omega ^{2}+2\sigma K)$.

\subsubsection{No group-velocity dispersion limit}

At the current state of the art in nanofabrication technology  the
interaction distance for parametric three-wave processes is strongly
limited by the losses of plasmonic oscillations  in metallic
nanostructures of LHM. The dispersion length is shorter than the
characteristic length of losses and therefore the dispersion term
can be omitted from  equations~(\ref{A51},\ref{A52}). We also
consider the case where the length of the LHM sample is shorter than
the characteristic length of losses.   Therefore, the effects of
group-velocity dispersion could be omitted as well.   The
corresponding system of equation takes the following form
\begin{equation}
ie_{1,\zeta }-e_{2}e_{1}^{\ast }=0,\quad ie_{2,\zeta }+i\delta
e_{2,\tau}-\Delta e_{2}+e_{1}^{2}/2=0.  \label{A5d1}
\end{equation}
The results of SHG computer modeling  using these equations were
presented in~\cite{786}. If $\Delta =0$, the system of equations
(\ref{A5d1}) is integrable using the inverse scattering
transform~\cite{R50} and also can be represented in bilinear Hirota
form~\cite{Hirot}.  Note, that in this particular case Hirota's
method is more convenient to analyze the nonstationary regime of SHG
in LHM.

\subsubsection{Non-collinear second harmonic generation}

In this section we  consider a process of  SHG, where pump waves and
the second harmonic wave  propagate in arbitrary directions.  In
this case the phase matching condition is determined by the beams'
orientations. The evolution of slowly varying optical pulse
envelopes is described by equations, which are similar to
(\ref{eq1}). The spatial derivative in this case must be replaced by
the corresponding directional derivatives  and the argument of the
exponential functions in the right-hand side must be replaced by
$\pm i(\mathbf{k}_{1}+\mathbf{k}_{2}-\mathbf{k}_{3})\cdot \mathbf{r
}$. The phase- matching condition in this case reads
\begin{equation}
\mathbf{k}_{1}+\mathbf{k}_{2}-\mathbf{k}_{3}=0  \label{eq7216}
\end{equation}
Let us assume that the second harmonic wave propagates along the
$z$-axis - $\mathbf{k}_{3}$ vector; both wave vectors -
$\mathbf{k}_{1}$ and $\mathbf{k}_{2}$ - of the pump waves are  in
the $xz$- plane;  so that $\mathbf{k}_{1}=k_{1}(\eta _{x},-\eta
_{z})$\ and $\mathbf{k}_{2}=k_{1}(-\eta _{x},-\eta _{z})$, where
$\eta _{x}=sin\theta$  and $\eta _{z}=cos\theta$, $\theta$ is an
angle between $\mathbf{k}_{1}$ and $\mathbf{k}_{3}$. The projection
of the vector equation (\ref{eq7216}) onto the $z$-axis defines the
phase-matching angle $\theta _{m}$: $n(2\omega )=n(\omega )\cos
\theta _{m}$. Let us assume that the group-velocity dispersion is of
no importance. The system of equations, describing the SHG, when
phase matching condition is satisfied, reads
\begin{eqnarray}
\left( -\eta _{z}\frac{\partial }{\partial z}+\eta
_{x}\frac{\partial }{
\partial x}+\frac{1}{v_{1}}\frac{\partial }{\partial t}\right) \mathcal{E}
_{1}^{(+)} &=&i\gamma _{1}\mathcal{E}_{1}^{(-)\ast }\mathcal{E}_{2}, \\
\left( -\eta _{z}\frac{\partial }{\partial z}-\eta
_{x}\frac{\partial }{
\partial x}+\frac{1}{v_{1}}\frac{\partial }{\partial t}\right) \mathcal{E}
_{1}^{(-)} &=&i\gamma _{1}\mathcal{E}_{1}^{(+)\ast }\mathcal{E}_{2},
\label{eqSHGn} \\
\left( \frac{\partial }{\partial z}+\frac{1}{v_{2}}\frac{\partial
}{\partial t}\right) \mathcal{E}_{2} &=&i\gamma
_{2}\mathcal{E}_{1}^{(+)}\mathcal{E} _{1}^{(-)},
\end{eqnarray}
where the pump wave is presented as%
\begin{eqnarray*}
E_{1}(t,\mathbf{r}) &=&\mathcal{E}_{1}^{(+)}(t,x,z)\exp \left[
-i\omega
t+ik_{1}(\eta _{x}x-\eta _{z}z)\right] + \\
&&+\mathcal{E}_{1}^{(-)}(t,x,z)\exp \left[ -i\omega t-ik_{1}(\eta
_{x}x+\eta _{z}z)\right] ,
\end{eqnarray*}
and the second harmonic wave is represented as%
\begin{equation*}
E_{2}(t,\mathbf{r})=\mathcal{E}_{2}(t,x,z)\exp \left[ -2i\omega
t+ik_{2}z) \right] .
\end{equation*}
System (\ref{eqSHGn}) can be transformed into a system of
dimensionless equations
\begin{equation}
ie_{1,\tau }+e_{3}^{\ast }e_{2}=0,~~ie_{2,\zeta
}+e_{1}e_{3}=0,~~ie_{3,\xi }+e_{1}^{\ast }e_{2}=0,  \label{eq7218}
\end{equation}
where $\zeta$, $\tau$ and $\xi$ are characteristic co-ordinates
defined as
\begin{eqnarray*}
\zeta &=&\eta _{z}v_{1}v_{2}\gamma _{1}\sqrt{\gamma
_{2}}A_{10}(v_{2}+\eta
_{z}v_{1})^{-1}\left( t-z/\eta _{z}v_{1}\right) , \\
\tau &=&2v_{1}v_{2}\gamma _{1}\sqrt{\gamma _{2}}A_{10}(v_{2}+\eta
_{z}v_{1})^{-1}\left( t-z/v_{2}+(v_{1}^{-1}+\eta
_{z}v_{2}^{-1})x/\eta
_{x}\right) , \\
\xi &=&2v_{1}v_{2}\gamma _{1}\sqrt{\gamma _{2}}A_{10}(v_{2}+\eta
_{z}v_{1})^{-1}\left( t-z/v_{2}-(v_{1}^{-1}+\eta
_{z}v_{2}^{-1})x/\eta _{x}\right).
\end{eqnarray*}
Normalized envelopes of the interacting waves are defined by the
expressions $\mathcal{E}_{1}^{(+)}=\sqrt{\gamma _{1}}A_{10}e_{1}$,
$\mathcal{E}_{2}=\sqrt{\gamma _{2}}A_{10}e_{2}$,
$\mathcal{E}_{1}^{(-)}=\sqrt{\gamma _{1}}A_{10}e_{3}$.

The solutions of these equations could be found either by the
inverse scattering method \cite{R57,Kaup80,Kaup81}, or  by using a
combination of simple algebraic manipulations and solution of
ordinary differential equations  \cite {R45}.

\subsection{Raman scattering process}

Let us consider optical waves propagating in an infinite medium and
scattered  by optical phonons. This process is known as Raman
scattering and differs from  Mandelschtam-Brillouin scattering of
optical waves occurs on  acoustic phonons. The phonon frequency
$\omega _{v} $ is much less than the carrier frequencies of optical
waves. The interaction of optical and vibration modes results in a
shift of the frequency of the optical mode. This interaction
generates two  new waves (Stokes waves) in addition to the pump wave
with carrier frequency $\omega _{L}$. The frequencies of these new
waves are   $\omega _{S}=\omega _{P}-\omega _{v}$ (Stokes wave), and
$\omega _{AS}=\omega _{L}+\omega _{v}$ ( anti-Stokes wave). In the
most cases, the intensity of the anti-Stokes wave is less than that
of the Stokes wave. Therefore, we consider only incident and Stokes
waves. Furthermore, we  consider stimulated Raman scattering when
the frequency $\omega _{P}$ lies in the positive refractive index
region and the frequency of the Stokes wave lies in negative
refractive index region.

The slowly varying envelopes of the incident and Stokes pulses is
governed by the system of the reduced Maxwell equations
\begin{eqnarray}
\left( \frac{\partial }{\partial z}+\frac{1}{v_{L}}\frac{\partial
}{\partial t}\right) \mathcal{E}_{L} &=&-\sigma
_{L}\mathcal{E}_{L}+i\frac{2\pi \omega _{L}^{2}\mu (\omega
_{P})}{c^{2}k_{L}}\mathcal{P}_{L}\exp (i\Delta kz),
\label{S21} \\
\left( -\frac{\partial }{\partial z}+\frac{1}{v_{s}}\frac{\partial
}{
\partial t}\right) \mathcal{E}_{S} &=&  \sigma _{S}\mathcal{E}_{S}+i\frac{
2\pi \omega _{S}^{2}\mu (\omega
_{S})}{c^{2}k_{S}}\mathcal{P}_{S}\exp (-i\Delta kz),  \label{S22}
\end{eqnarray}
The linear losses are taken into account by introducing extra terms
with the coefficients $\sigma _{L,S}$.

We define polarizations $\mathcal{P}_{S,L}$ in the Raman medium  in
accordance  with Placzek's classical  model where  molecules  are
represented by  harmonic oscillators. The polarization of this
medium is represented  as $P=n_{A}\alpha (Q)E$, where $n_{A}$ is the
concentration of the molecules, $\alpha (Q)$ is the molecular
polarizability, and $Q$ is the vibrational co-ordinate of a molecule
determining the magnitude of deflection from equilibrium. In the
case of small vibrations the molecular polarizability is expressed
by the first two terms of its Taylor series, i.e.,
\begin{equation*}
\alpha (Q)\approx \alpha _{0}+(\partial \alpha /\partial
Q)_{0}Q\equiv \alpha _{0}+\alpha _{D}Q.
\end{equation*}
Hence, the non-linear polarization is $P_{NL}=n_{A}\alpha _{D}QE$.
In the slowly varying envelope approximation $Q$ is represented as
follows
\begin{equation*}
Q=u(t,z)\exp \left( -i\omega _{v}t+ik_{v}z\right) +u^{\ast
}(t,z)\exp \left( i\omega _{v}t-ik_{v}z\right) ,
\end{equation*}
where  $u(t,z)$ is determined by the  equation
\begin{equation}
u_{,t}-i(\omega _{v}-\omega _{S})u+\sigma _{v}u=i\left( \alpha
_{D}/2m\omega _{v}\right) \mathcal{E}_{L}^{\ast }\mathcal{E}_{S}\exp
(i\Delta kz), \label{S23}
\end{equation}
Here $m$ is the effective molecule mass. The coefficient $\sigma
_{v}$\ takes into account linear losses. Taking into account that
the phonon wave number is much smaller than the photon wave numbers,
the mismatch value can be taken to be: $\Delta k\approx
k_{S}-k_{L}$. We assume that spectral half-width of the ultrashort
optical pulse is much smaller than the phonon frequency  $\Delta
\omega _{p}\ll $\ $\omega _{v}$ and express  the nonlinear slowly
varying amplitudes of polarization as $\mathcal{P}_{L}=n_{A}\alpha
_{D}u^{\ast }\mathcal{E}_{S}$ and $\mathcal{P}_{S}=n_{A}\alpha _{D}u\mathcal{%
E}_{L}$.

To develop a theory of  Raman scattering we can use the system of
equations (\ref{S21}-\ref{S23}). Note that in the case we consider
the Stokes wave and the laser wave are propagating in opposite
directions.

It is convenient to introduce a co-moving system of coordinates
$\zeta =z/L$, $\tau =(t-z/V_{0})/t_{p}$  with velocity $V_{0}$
defined by as $V_{0}^{-1}=\left( v_{L}^{-1}-v_{S}^{-1}\right) /2$,
and normalized functions $u=q/(n_{A}\alpha _{D})$,
$\mathcal{E}_{L}=\sqrt{\gamma _{L}} A_{0}e_{p}\exp (i\Delta kz)$,
$\mathcal{E}_{S}=\sqrt{\gamma _{S}}A_{0}e_{s}$, where
\begin{eqnarray*}
\gamma _{S} &=&(2\pi \omega _{S}c^{-1}\sqrt{\mu (\omega
_{S})/\varepsilon (\omega _{S})},\gamma _{L}=(2\pi \omega
_{P}c^{-1}\sqrt{\mu (\omega
_{L})/\varepsilon (\omega _{L})}, \\
L^{-1} &=&\sqrt{\gamma _{S}\gamma _{L}},\quad A_{0}^{-2}=\left(
n_{A}\alpha _{D}^{2}\sqrt{\gamma _{S}\gamma _{L}}\right) /2m\omega
_{v}.
\end{eqnarray*}
The dimensionless system of equations takes the form
\begin{eqnarray}
ie_{p,\zeta }+i\delta e_{p,\tau }-\Delta e_{p}+i\Gamma
_{p}e_{p}+q^{\ast
}e_{s} &=&0,  \label{S81}\nonumber \\
ie_{s,\zeta }-i\delta e_{s,\tau }-i\Gamma _{s}e_{s}-qe_{p} &=&0,
\label{S82}
\\
iq_{,\tau }+\vartheta q+i\Gamma _{v}q+e_{s}e_{p}^{\ast } &=&0,
\label{S83}\nonumber
\end{eqnarray}
where $\delta =Lt_{p}^{-1}(v_{1}^{-1}+v_{2}^{-1})/2$, $\vartheta
=t_{p}(\omega _{v}-\omega _{S})$, $\Gamma _{p}=\sigma _{L}L$,
$\Gamma _{s}=\sigma _{S}L,\Gamma _{v}=\sigma _{v}L$. Therefore the
system of equations describing Raman scattering is represented as a
three wave interaction model.

In the case of strong vibrational damping  ($\Gamma _{p}\gg1$), it
follows from the third equation of (\ref{S82}) that  $q\approx
i\Gamma _{v}^{-1}e_{s}e_{p}^{\ast }$. Substitution of this formula
into the two first equations of  (\ref{S82})\ leads to  equations
describing a Raman amplifier in LHM
\begin{eqnarray}
ie_{p,\zeta }+i\delta e_{p,\tau }-\Delta e_{p}+i\Gamma
_{p}e_{p}-i\Gamma
_{v}^{-1}|e_{s}|^{2}e_{p} &=&0,  \nonumber \\
ie_{s,\zeta }-i\delta e_{s,\tau }-i\Gamma _{s}e_{s}-i\Gamma
_{v}^{-1}|e_{p}|^{2}e_{s} &=&0.  \nonumber
\end{eqnarray}
Note, that choosing $\Gamma _{p}=\Gamma _{s}=0$\ and $\Gamma _{v}\gg
1$, this system of equations can be used to study the phenomena  of
Raman spike generation in left handed materials. This phenomena is
well known in conventional right-handed
materials~\cite{DWC83,WCDS85}.

\subsection{Third Harmonic Generation in NRI medium}

Third harmonic generation (THG) is associated with four wave
interaction of the type $\omega _{1}+\omega _{1}+\omega
_{1}\rightarrow \omega _{3}=3\omega _{1}$. This process is described
by the following system of equation
\begin{eqnarray}
\left( \hat{k}_{1}\frac{\partial }{\partial
z}+\frac{1}{v_{1}}\frac{\partial }{\partial t}\right)
\mathcal{E}_{1}-\frac{D_{1}}{2}\frac{\partial ^{2}
\mathcal{E}_{1}}{\partial t^{2}} &=&i\frac{2\pi \omega _{1}^{2}\mu
(\omega
_{1})}{c^{2}k_{1}}\tilde{P}_{NL}(\omega _{1})\exp (-ik_{1}z)\label{AT1} \\
\left( \hat{k}_{3}\frac{\partial }{\partial
z}+\frac{1}{v_{3}}\frac{\partial }{\partial t}\right)
\mathcal{E}_{3}-\frac{D_{3}}{2}\frac{\partial ^{2}
\mathcal{E}_{3}}{\partial t^{2}} &=&i\frac{2\pi \omega _{3}^{2}\mu
(\omega _{3})}{c^{2}k_{3}}\tilde{P}_{NL}(\omega _{3})\exp
(-ik_{3}z)\label{AT2}
\end{eqnarray}
where
\begin{eqnarray*}
\tilde{P}_{NL}(\omega _{1}) &=&\chi ^{(3)}(\omega _{1};\omega
_{3},-\omega _{1},-\omega _{1})\mathcal{E}_{3}\mathcal{E}_{1}^{\ast
2}\exp
[iz(k_{3}-2k_{1})]+ \\
&&+\chi ^{(3)}(\omega _{1};\omega _{1},-\omega _{1},\omega
_{1})|\mathcal{E} _{1}|^{2}\mathcal{E}_{1}+\chi ^{(3)}(\omega
_{1};\omega _{1},-\omega
_{3},\omega _{3})|\mathcal{E}_{3}|^{2}\mathcal{E}_{1}, \\
\tilde{P}_{NL}(\omega _{2}) &=&\chi ^{(3)}(\omega _{3};\omega
_{1},\omega
_{1},\omega _{1})\mathcal{E}_{1}^{3}\exp [iz3k_{1})]+ \\
&&+\chi ^{(3)}(\omega _{3};\omega _{3},-\omega _{3},\omega
_{3})|\mathcal{E} _{3}|^{2}\mathcal{E}_{3}+\chi ^{(3)}(\omega
_{3};\omega _{3},-\omega _{1},\omega
_{1})|\mathcal{E}_{1}|^{2}\mathcal{E}_{3}.
\end{eqnarray*}
In contrast to SHG, the process of transformation of one wave to
another in this case is accompanied by the two additional effects of
self-phase and cross-phase modulation.

Let us consider the THG where the frequency of the fundamental wave
(pump wave) $\omega _{1}$ is located in the NRI spectral region, and
the third harmonic frequency $\omega _{3}$ is located in the
positive refractive index spectral region. Let the self-modulation
and cross-modulation effects be ignored. In this case THG equations
after normalization read as
\begin{equation}
\begin{array}{lcl}
ie_{1,\zeta }+(\sigma /2)e_{1,\tau \tau }-e_{2}e_{1}^{\ast 2} & = & 0, \\
ie_{3,\zeta }+i\delta e_{3,\tau }-(\beta /2)e_{3,\tau \tau }-\Delta
e_{3}+e_{1}^{3} & = & 0,
\end{array}
\label{eqT5}
\end{equation}
where both parameters and normalized variables were introduced along
similar lines to this use for SHG.

\subsubsection{Third harmonic generation. CW-limit}

First we consider   third harmonic generation in the case of
continuous waves. The system of equations describing THG in the
continuous wave limit takes the following form
\begin{equation}
ie_{1,\zeta }-e_{2}e_{1}^{\ast 2}=0,\quad ie_{3,\zeta }-\Delta
e_{3}+e_{1}^{3}=0.  \label{T6}
\end{equation}
There are two types of boundary conditions: (a) $\left\vert
e_{1}(\zeta =0)\right\vert =a_{0}$,$\left\vert e_{3}(\zeta
\rightarrow \infty )\right\vert =0$, i.e., infinite medium, and (b)
$\left\vert e_{1}(\zeta =0)\right\vert =a_{0}$,$\left\vert
e_{3}(\zeta =l)\right\vert =0$ , i.e., THG in the plate having width
equal to $l$.

The Manley-Rowe relation following from (\ref{T6}) is as follows
\begin{equation}
\left\vert e_{1}\right\vert ^{2}-\left\vert e_{3}\right\vert
^{2}=c_{0}^{2}= \mathrm{const.}  \label{T8}
\end{equation}
Let consider the THG at phase mismatch $\Delta =0$. The solution of
the system (\ref{T6}) for infinite medium, can be written as
\begin{equation}
\left\vert e_{1}(\zeta )\right\vert =\left\vert e_{3}(\zeta
)\right\vert = \frac{a_{0}}{\sqrt{1+2a_{0}^{2}\zeta }}.  \label{T11}
\end{equation}
Hence, in this special case there is complete transformation of the
incident wave to a third harmonic wave which propagates in the
opposite direction. The solution of (\ref{T6}) in the more realistic
case, where the  size of the nonlinear dielectric  sample $l$ is
finite,  reads as
\begin{equation}
\left\vert e_{1}(\zeta )\right\vert ^{2}=\frac{c_{0}^{2}\xi
^{2}}{1-\xi ^{2}} ,\quad \left\vert e_{3}(\zeta )\right\vert
^{2}=\frac{c_{0}^{2}\xi ^{2}}{ 1-\xi ^{2}}, ~~\xi =c_{0}^{2}(\zeta
-l). \label{T16}
\end{equation}

The equation for the unknown constant $c_{0}^{2}$ follows from the
Manley-Rowe relation:
\begin{equation}
a_{0}^{2}(1-l^{2}c_{0}^{4})=c_{0}^{2}.\nonumber
\end{equation}
Note that in contrast to the case of SHG there is only one solution
of this equation which has physical meaning.  In the case of SHG the
corresponding equation  reads $la_{0}\sin c_{0}l=\sqrt{2}c_{0}l$ and
in general can have several solution~\cite{Popov,Popov1}. Like in
the case of SHG, this Manley-Rowe relation follows from the new type
of phase-matching condition referred in~\cite{Popov,Popov1}  as
\textquotedblleft backward phase-matching\textquotedblright\ . This
type of phase-matching is a fundamental feature of  wave interaction
in NRI medium.

\subsubsection{Solitary wave solutions of third harmonic generation equations}

A solitary wave  solution (coupled state of two waves with carrier
frequencies $\omega _{1}$ and $ \omega _{3}$)  can be found as in
the case of SHG. Let us introduce the real variables for interacting
waves $e_{1}=a\exp (i\varphi _{1})$ and $e_{3}=b\exp (i\varphi
_{3})$, where $\varphi _{1}=K\zeta +\Omega \tau $, $\varphi
_{3}=3K\zeta +3\Omega \tau $, $a=a(\tau -\zeta /V)$, $ b=b(\tau
-\zeta /V)$. The system of the equations (\ref{eqT5}) reduces to a
single equation for the real amplitude $a$, if the $V^{-1}=\sigma
\Omega $, frequency $\Omega =\delta /(\sigma +3\beta )$ , and wave
number is determined by formula
\begin{equation*}
K=\sigma \left( 4\beta \Omega ^{2}-3\delta \Omega -\Delta \right)
/(3\sigma +\beta ).
\end{equation*}
Solving this equation we found the field amplitudes
\begin{equation*}
\left\vert e_{1}(\tau ,\zeta )\right\vert =\frac{\sqrt{p/f}}{\cosh
[p^{1/2}(\tau -\zeta /V-y_{0})]},\quad \left\vert e_{3}(\tau ,\zeta
)\right\vert =\frac{\sqrt{pf}}{\cosh [p^{1/2}(\tau -\zeta
/V-y_{0})]},
\end{equation*}
where $y_{0}$ is a constant of integration, $f^{2}=\sigma /\beta $\
and $ p=2\sigma K+\Omega ^{2}$. For $\sigma =\beta =-1$\ we can find
$\Omega =-\delta /4$, $K=(\delta ^{2}-\Delta )/8$ and $p=\left(
4\Delta -3\delta ^{2}\right) /16$. The velocity of this soliton is
defined as $\tau -\zeta /V=(t-z/V_{s})/t_{p}$ and is equal to
$V_{s}=4v_{1}v_{3}/(v_{1}-3v_{3})$.

Note that  the system of equations  (\ref{eqT5}) has periodic
cnoidal solutions. These solutions describe  nonlinear periodic
waves in a medium with cubic nonlinearity.

In the  large-mismatch limit of THG, the system of
equations~(\ref{eqT5}) transforms into  the quintic nonlinear
Schr\H{o}dinger equation:
\begin{equation*}
ie_{1,\zeta }+(\sigma /2)e_{1,\tau \tau }-(1/2\Delta
)|e_{1}|^{4}e_{1}=0.
\end{equation*}
This equation also has  solitary  pulse like and  periodic cnoidal
wave  solutions.

\subsubsection{Phase interaction of two waves}

There are situations where the process of conversion of waves with
different frequencies can be ignored.  For example, such a situation
occurs when the phase difference of the two interacting waves is
rapidly changing in space: $\Delta k \times L_{d}\gg1$, here $L_{d}$
is the dispersion length. Let us consider  propagation of two waves
in a medium with cubic nonlinearity, where the frequency of one wave
is in the NRI region, and  the frequency of the other wave is in the
PRI region. The corresponding equations in dimensionless form, which
follow from~(\ref{AT1},\ref{AT2}), read as
\begin{equation}
\begin{array}{lcl}
ie_{1,\zeta }+(\sigma /2)e_{1,\tau \tau }-\left( |e_{1}|^{2}+\mu
|e_{3}|^{2}\right) e_{1} & = & 0, \\
ie_{3,\zeta }+i\delta e_{3,\tau }-(\beta /2)e_{3,\tau \tau }+\left(
|e_{3}|^{2}+\mu |e_{1}|^{2}\right) e_{3} & = & 0,
\end{array}
\label{TMan}
\end{equation}
This system represents  a generalization of the Manakov's equations
\cite{Manakov}. The solitary wave solutions of these equations can
be found by standard methods.

\section{Periodic structures}

In recent years, much attention has been drawn to various artificial
structures, which are one- two- or three-dimensional periodic
dielectric (or metallic) structures, frequently referred to as
photonic crystals. The spectrum of  electromagnetic waves in these
structures possess bandgaps (the Bragg gaps) \cite{Joan}. Sometimes
the photonic crystals are called photonic band-gap materials (PBG).
In addition to a conventional PBG, a one-dimensional periodic
structure consisting of alternating layers of PRI and NRI medium
represents a new type of photonic band-gap material, a so-called
zero refractive index PBG \cite{Li03}. There we consider only the
resonant Bragg grating and nonlinear optical waveguide array as
simple examples of nonlinear periodic medium.

\subsection{Resonant Bragg grating}

In the simplest case the \emph{resonant Bragg grating} \cite
{Ma1,Ma2,Ko1,Ko2,KKOM} \ consists of a linear homogeneous dielectric
medium containing an array of thin films with resonant atoms or
molecules. The distance between successive films is $a$, and the
thickness of the film $ l_{f}$ is much less than the wavelength of
the electromagnetic wave propagating through such a structure. The
interaction of ultra-short pulses and films embedded with two-level
atoms have been studied by Mantsyzov et al. \cite{Ma1,Ma2}  in the
framework of the two-wave reduced Maxwell-Bloch model and by
Kozhekin \cite{Ko1,Ko2}. This work demonstrated the existence of the
$ 2\pi $-pulse of self-induced transparency in such structures. It
was also found \cite{R8} that bright as well as dark solitons can
exist in the prohibited spectral gap, and that bright solitons can
have arbitrary pulse area. If the density of two-level atoms is very
high, then the near-dipole-dipole interaction is noticeable and
should be accounted for in the mathematical model. The effect of
dipole-dipole interaction on the existence of gap solitons in a
resonant Bragg grating was studied in Ref. \cite{R10}. The unusual
solution known as a $\emph{zoomeron}$ was discovered and
investigated recently in the context of the resonant Bragg grating
\cite {Mant05}. A zoomeron is a localized pulse similar to an
optical soliton, except that its velocity oscillates about some mean
value.

Recent advances in nanofabrication have allowed the creation of
nanocomposite materials which have the ability to sustain nonlinear
plasmonic oscillations. These materials have metallic nanoparticles
embedded in them \cite{R97,DBNS04,HRF86}. In \cite{MGK07} a
dielectric into which thin films containing metallic nanoparticles
have been inserted was considered. These thin films are spaced
periodically along the length of the dielectric so that the Bragg
prohibited spectral gap is centered at the plasmonic resonance
frequency of the nanoparticles. A set of discrete equations
describes the propagation of light through a medium of alternating
layers of linear dielectric and thin film. However, the physical
parameters of the system make the slowly-varying envelope
approximation appropriate. Following reviews \cite{SS94,K99,KA05},
we employ such an approximation to derive envelope equations for two
counter-propagating electric fields which are coupled through the
medium polarization. The electric field can be described by
\begin{equation*}
E(x,t)=\left\{
\mathcal{A}(x,t)e^{iq_{0}x}+\mathcal{B}(x,t)e^{-iq_{0}x} \right\}
e^{-i\omega _{0}t}.
\end{equation*}
\noindent Here $\omega _{0}$ is the carrier frequency, and
$q_{0}=\omega _{0} \sqrt{\varepsilon }/c$ is the wavevector in an
optical medium with electric permittivity $\varepsilon $. The
long-wave and slowly varying envelope of the pulses approximations
result in the system of coupled wave equations.
\begin{eqnarray}
i\left( \mathcal{A}_{,x}+v_{g}^{-1}\mathcal{A}_{,t}\right) +\Delta
q_{0} \mathcal{A} &=&-(2\pi \omega _{0}/c\sqrt{\varepsilon
})\mathcal{P},
\label{GM1} \\
i\left( \mathcal{B}_{,x}-v_{g\mathcal{B},t}^{-1}\right) -\Delta
q_{0} \mathcal{B} &=&+(2\pi \omega _{0}/c\sqrt{\varepsilon
})\mathcal{P}, \label{GM2}
\end{eqnarray}
where $\Delta q_{0}=q_{0}-2\pi /a$ is the mismatch between the
carrier wavenumber and the Bragg resonant wavenumber. Description of
the slowly varying polarization is based on the anharmonic
oscillator model (\ref{eqD83}):
\begin{equation*}
\mathcal{P}_{,t}-i(\omega _{0}-\omega
_{d})\mathcal{P-}(3ig_{p}/2\omega
_{0})|\mathcal{P}|^{2}\mathcal{P}=i(\omega _{p}^{2}/8\pi \omega
_{0}) \mathcal{E}_{f},
\end{equation*}
where $\mathcal{E}_{int}$ is the electric field interacting with
metallic nanoparticles. In our case we have $\mathcal{E}_{int}=
\mathcal{A}+\mathcal{B}$. Note, that if anharmonicity of oscillators
is neglected  ($g_{p}=0$) then  the system of equations transforms
into the well known  Lorentz model.   In the general case the system
of resulting equations is the \emph{two-wave Maxwell-Duffing
equations}. Using rescaling
\begin{eqnarray*}
f_{s} &=&-(\mathcal{A}+\mathcal{B})A_{0}^{-1}e^{-i\delta \tau
},\quad \quad
f_{a}=(\mathcal{A}-\mathcal{B})A_{0}^{-1}e^{-i\delta \tau }, \\
q &=&(4\pi \omega _{0}/[\sqrt{\varepsilon }\omega
_{p}A_{0}])e^{-i\delta \tau }\mathcal{P},~~\zeta =(\omega
_{p}/2c)x,~~\tau =t/t_{0}.
\end{eqnarray*}
these equations can be represented in dimensionless form:
\begin{equation}
\begin{array}{ccc}
f_{a,\zeta \zeta }-f_{a,\tau \tau } & = & 2iq_{,\zeta } \\
f_{s,\zeta \zeta }-f_{s,\tau \tau } & = & 2iq_{,\tau } \\
iq_{,\tau }+(\Delta -\delta )q+\mu |q|^{2}q & = & f_{s}.
\end{array}
\label{GM5}
\end{equation}
Here $t_{0}=2\sqrt{\varepsilon }/\omega _{p}$ is an inverse plasma
frequency,  $A_{0}$ is characteristic amplitude of
counter-propagating fields, $\mu =(3g_{p}\sqrt{\varepsilon }/\omega
_{0}\omega _{p})(\sqrt{\varepsilon }\omega _{p}/4\pi \omega
_{0})^{2}A_{0}^{2}$ is a dimensionless coefficient of anharmonicity,
$\delta =2\Delta q_{0}(c/\omega _{p})$ is the dimensionless mismatch
coefficient, $\Delta =2\sqrt{\varepsilon }(\omega _{d}-\omega
_{0})/\omega _{p}$ is dimensionless detuning of a nanoparticle's
resonance frequency from the field's carrier frequency.

The exact solitary wave solutions of the system (\ref{GM5}) have
been found in~\cite{MGK07} It was shown that, in contrast to
conventional $2\pi $-pulses, they have nonlinear phase. The
stability of these solutions is sensitive to the phase
perturbations. It was  also demonstrated that   the outcomes of
pulse collisions are highly dependent on relative phase.

\subsection{Nonlinear optical waveguide array}

A simple example of a waveguide structure is the \emph{directional
coupler}. The directional coupler is represented by two parallel
waveguides.  The separation distance between these waveguides so
small that guided light is leaking from one waveguide to another.
Sometimes these waveguides are described as tunnel-connected
\cite{Integ}. If the sign of the refractive index in both waveguides
is positive, then a beam launched into one of the waveguides
produces  waves in both waveguides propagating in the same
directions. If  the index of refraction of one of the waveguides is
negative, then a beam launched into one of the waveguides produces
two beams in these waveguides propagating in opposite
directions~\cite{AE05}. The theory of this \emph{anti-directional
coupler} is similar to the description of SHG under the undepleted
pump approximation. Hence, the anti-directional coupler acts as
distributed mirror \cite{Integ}. If the waveguides are made from
non-linear dielectric material or if they are embedded into a
non-linear medium, then the coupling performance, in addition to the
coupler's geometry and material property, depend  on the wave input
power. If input power is above a certain threshold the
antidirectional coupling property can be transformed to become
unidirectional.

Nonlinear optical waveguide arrays (NOWA) are a natural
generalization of nonlinear couplers. NOWA with a positive
refractive index have many useful applications and are well studied
in the literature (see for example~\cite{Ch,Sch,Darm97}). Materials
with negative index of refraction offer  new types of array
structures which can be combined in three groups.The first group
corresponds to an array of alternating waveguides with different
signs of refractive index. The second group corresponds to the
situation when one planar array containing identical waveguides with
positive sign of refractive index is attached to another planar
array of waveguides with negative sign of refractive index.  The
third group corresponds to a situation where an array of identical
waveguides contains a single waveguide with opposite sign.

To develop a theory of such waveguides it is natural to utilize
existing formalism based on nearest-neighbor coupling.  The electric
field of an optical wave propagating in NOWA in the positive $z$
direction can be represented as follows
\begin{equation*}
E(x,y,z,t)=\sum\limits_{J=-\infty }^{J=+\infty
}\sum\limits_{m}A_{m}^{(J)}(z,t)\Psi _{m}^{(J)}(x,y)\exp \left(
-i\omega _{0}t+i\beta _{m}^{(J)}z\right) .
\end{equation*}
The mode function for a particular $m$-th mode of channel $J$ is
denoted by $\Psi _{m}^{(J)}(x,y)$, where $J=0,\pm 1,\pm 2,\ldots $,
and $ A_{m}^{(J)}(z,t)$ is a slowly varying envelope of the electric
field corresponding to this mode. Parameters $\beta_{m}^{(J)}$ are
propagation constants.  Omitting the details we can write the
general equations which are governed by normalized envelopes
$q_{m}^{(J)}=A_{m}^{(J)}(z,t)/A_{0}$ :
\begin{equation}
\begin{array}{lcl}
i\hat{k}_{J}q_{J,z}+iv_{g}^{-1}q_{J,t}-\sigma
_{J}q_{J,tt}+K_{12}\left(
q_{J-1}+q_{J+1}\right) & + &  \\
+\left( 2\pi \omega _{0}^{2}\mu (\omega _{0})/c^{2}\beta
_{{}}^{(J)}\right) \chi _{eff}A_{0}^{2}|q_{J}|^{2}q_{J} & = & 0.
\end{array}
\label{NOWA1}
\end{equation}
Here we assume that the group velocities and the propagation
constants are equal for all channels and only resonant coupled modes
are essential for our further consideration ($\sigma _{J}=\sigma
_{gm}^{(J)}$ and $v_{g}=v_{gm}^{(J)}$).  The coefficient $K_{12}$ is
the coupling constant between neighboring waveguides, and $\chi
_{eff}$ is the effective non-linear susceptibility. The amplitude
$A_{0}$ is chosen to normalize the nonlinear coefficient of
self-interaction to one.

Using equations~(\ref{NOWA1}) and taking into account that
$(\hat{k}_{J}=+1)$ and $(\hat{k}_{J}=-1)$ corresponds to PRI and NRI
cases respectively, we can present mathematical models describing
the three groups of arrays listed above.

\subsubsection{Alternated nonlinear optical waveguide array}

Let us suppose  that waveguides marked as $J=2n$ and $J=2n+1$ have
PRI and NRI properties respectively. The system of equations
describing electromagnetic wave propagation in this structure reads
\begin{equation*}
\begin{array}{l}
\displaystyle iq_{2n,z} + iv_{g}^{-1}q_{2n,t}-\sigma _{2n}q_{2n,tt}
+K_{12}\left( q_{2n-1} + q_{2n+1}\right) +|q_{2n}|^{2}q_{2n} = 0,\\
\displaystyle iq_{2n+1,z} - iv_{g}^{-1}q_{2n+1,t}+\sigma
_{2n+1}q_{2n+1,tt} -K_{12}\left( q_{2n} + q_{2n+2}\right)
-|q_{2n+1}|^{2}q_{2n+1} = 0.
\end{array}
\end{equation*}
In the linear limit these equations correspond to the model of a
harmonic lattice with alternated sign for the coupling. In this case
the spectrum of the linear waves consist of two dispersion branches.
This means the alternated linear optical waveguide array acts as a
gap medium. Therefore, in the nonlinear case we can expect the
existence the solitary waves.

\subsubsection{Interface of the nonlinear optical waveguide arrays}

Let suppose that waveguides marked as$n\geq 0$ and $n<0$ have PRI
and NRI properties respectively.  In this case propagation of
electromagnetic waves is governed by the following system of
equations
\begin{eqnarray*}
iq_{n,z}+iv_{g}^{-1}q_{n,t}-\sigma _{n}q_{n,tt}+K_{12}\left(
q_{n-1}+q_{n+1}\right) +|q_{n}|^{2}q_{n} &=&0,\quad n\geq 0, \\
iq_{n,z}-iv_{g}^{-1}q_{n,t}+\sigma _{n}q_{n,tt}-K_{12}\left(
q_{n-1}+q_{n+1}\right) -|q_{n}|^{2}q_{n} &=&0,\quad n<0.
\end{eqnarray*}
In the  linear approximation, due to the opposite directionality of
phase and group velocities in NRI materials, one can expect
existence of surface waves with vortex structures. Each type of
nonlinear infinite array supports solitary wave solutions. One can
expect that the boundary effect at the interface of two types of
arrays will be revealed through  pinning of solitary wave solutions
in a way analogous to the behavior of light near defects in a
photonic crystal.

\subsubsection{Nonlinear optical waveguide array with a defect}

Let us assume that the PRI type array of waveguides contains one
waveguide of NRI type. This waveguide, marked $n=0$, can be viewed
as a defect in a one dimensional lattice. Propagation of an
electromagnetic wave in such an array is governed by the equation:
\begin{eqnarray*}
iq_{n,z}+iv_{g}^{-1}q_{n,t}-\sigma _{n}q_{n,tt}+K_{12}\left(
q_{n-1}+q_{n+1}\right) +|q_{n}|^{2}q_{n} &=&0,\quad n>0, \\
iq_{0,z}-iv_{g}^{-1}q_{0,t}+\sigma _{0}q_{0,tt}-K_{12}\left(
q_{-1}+q_{+1}\right) -|q_{0}|^{2}q_{0} &=&0,\quad n=0, \\
iq_{n,z}+iv_{g}^{-1}q_{n,t}-\sigma _{n}q_{n,tt}+K_{12}\left(
q_{n-1}+q_{n+1}\right) +|q_{n}|^{2}q_{n} &=&0,\quad n<0.
\end{eqnarray*}
Presence of a ``defect'' could be a reason for a pinning phenomena,
where the electromagnetic wave is trapped in this waveguide. One can
expect that increase of the localized field energy will destroy
localization.

\section{Conclusion}

We considered  new features of classical phenomena of nonlinear
optics in optical metamaterials including materials with negative
refractive index. In addition to already obtained results in this
field, we discussed a number of problems of potential importance for
the nonlinear optics of metamaterials.

Optical transparency of metamaterials is a fundamental assumption in
our considerations. The current state of the art in nanofabrication
technology is capable of delivering only thin film metamaterials
with a relatively large level of losses. We expect that further
development in nanofabrication will overcome problems of
metamaterial losses. Our expectations are based on the history of
optical fiber technology development  in the second part of the last
century.

\section*{Acknowledgment}

We would like to express our gratitude to J-G.~Caputo, R.~Indik,
N.M.~Litchinitser, B.I.~Mantsyzov,  and  A.A.~Zabolotskii    for
enlightening helpful discussions.  A.I. Maimistov thanks the
Department of Mathematics of University  of Arizona  and Laboratoire
de Math\'{e}matiques, INSA de Rouen for the support and hospitality
while he looked into the problems under consideration. This study
was supported in part by the Russian Foundation for Basic Research
(grant no. 06-02-16406), by NSF (grant DMS-0509589), ARO-MURI award
50342-PH-MUR  and State of Arizona (Proposition 301).

\end{document}